\documentclass[aps,floatfix,superscriptaddress,preprint,preprintnumbers]
{revtex4}

\usepackage{titlesec}
\usepackage{amssymb,amsmath,amsfonts,latexsym,graphicx,epsfig,bm}
\usepackage{epstopdf}
\usepackage{booktabs}
\titleformat{\section}{\large\bfseries}{\thesection}{1em}{}

\newcommand{\bea}{\begin{eqnarray}}
\newcommand{\ena}{\end{eqnarray}}
\newcommand{\nn}{\nonumber\\}
\newcommand{\be}{\begin{equation}}
\newcommand{\en}{\end{equation}}

\newcommand{\ed}{\end{document}}

\newcommand{\slp}{p\kern-5pt/}

\begin{document}

\title{\boldmath{$D^\ast$} Polarization as an Additional Constraint on New Physics in the \boldmath{$b\to c\tau\bar\nu_\tau$} Transition}

\author {M. A. Ivanov}
\email{ivanovm@theor.jinr.ru}
\affiliation{
Bogoliubov Laboratory of Theoretical Physics,
Joint Institute for Nuclear Research,
141980 Dubna, Russia}

\author{J. G. K\"{o}rner}
\email{jukoerne@uni-mainz.de}
\affiliation{PRISMA Cluster of Excellence, Institut f\"{u}r Physik, 
Johannes Gutenberg-Universit\"{a}t, 
D-55099 Mainz, Germany}

\author{P. Santorelli}
\email{Pietro.Santorelli@na.infn.it}
\affiliation{Dipartimento di Fisica, Universit\`{a} di Napoli Federico II, Complesso Universitario di Monte S. Angelo, Via Cintia, Edificio 6, 80126 Napoli, Italy}
\affiliation{Istituto Nazionale di Fisica Nucleare, Sezione di Napoli, 80126 Napoli, Italy}

\author{C. T. Tran}
\email{tranchienthang1347@gmail.com}
\affiliation{Dipartimento di Fisica, Universit\`{a} di Napoli Federico II, Complesso Universitario di Monte S. Angelo, Via Cintia, Edificio 6, 80126 Napoli, Italy}
\affiliation{Istituto Nazionale di Fisica Nucleare, Sezione di Napoli, 80126 Napoli, Italy}

\preprint{MITP/20-034}

\date{\today}

\begin{abstract}
Measurements of the branching fractions of the semileptonic decays $B\to D^{(*)}\tau\bar\nu_\tau$ and $B_c\to J/\psi\tau\bar\nu_\tau$ systematically exceed the Standard Model (SM) predictions, pointing to possible signals of new physics that can violate lepton flavor universality. The unknown origin of new physics realized in these channels can be probed using a general effective Hamiltonian constructed from four-fermion operators and the corresponding Wilson coefficients. Previously, constraints on these Wilson coefficients were obtained mainly from the experimental data for the branching fractions. Meanwhile, polarization observables were only theoretically studied. The situation has changed with more experimental data having become available, particularly those regarding the polarization of the tau and the $D^*$ meson. In this study, we discuss the implications of the new data on the overall picture. We then include them in an updated fit of the Wilson coefficients using all hadronic form factors from our covariant constituent quark model. The use of our form factors provides an analysis independent of those in the literature. 
Several new-physics scenarios are studied with the corresponding theoretical predictions provided, which are useful for future experimental studies.
%%---Beginning of addition
{In particular, we find that under the one-dominant-operator assumption, no operator survives at $1\sigma$. Moreover, the scalar operators ${\cal O}_{S_L}$ and ${\cal O}_{S_R}$ are ruled out at $2\sigma$ if one uses the constraint ${\cal B}(B_c\to\tau\nu_\tau)\leq 10\%$, while the more relaxed constraint ${\cal B}(B_c\to\tau\nu_\tau)\leq 30\%$ still allows these operators  at $2\sigma$, but only minimally.
The inclusion of the new data for the $D^*$ polarization fraction $F_L^{D^*}$ reduces the likelihood of the right-handed vector operator ${\cal O}_{V_R}$ and significantly constrains the tensor operator ${\cal O}_{T_L}$. Specifically, the $F_L^{D^*}$ alone rules out ${\cal O}_{T_L}$ at $1\sigma$. Finally, we show that the longitudinal polarization $P_L^\tau$ of the tau in the decays $B\to D^*\tau\bar\nu_\tau$ and $B_c\to J/\psi\tau\bar\nu_\tau$ is extremely sensitive to the tensor operator. Within the $2\sigma$ allowed region, the best-fit value $T_L=0.04+i0.17$ predicts $P_L^\tau(D^*)=-0.33$ and $P_L^\tau(J/\psi)=-0.34$, which are at about 33\% larger than the SM prediction $P_L^\tau(D^*)=-0.50$ and $P_L^\tau(J/\psi)=-0.51$.}
%%---End of addition
\end{abstract}

%\pacs{}

%\keywords{semileptonic decay, $B$ meson, lepton flavor universality, beyond Standard Model, form factors, covariant constituent quark model, branching fraction, polarization}

\maketitle

\section{Introduction}
\label{sec:intro} 
The Standard Model (SM) of elementary particles has been tested in numerous high-precision experiments, showing its uniquely powerful predicting ability in a wide range of physical processes. However, the lack of answers to fundamental questions, such as the problems of hierarchy, dark matter, neutrino mass, etc., implies that the SM can well be a low-energy effective theory of  a more fundamental one. Therefore, the search for New Physics (NP) beyond the SM is one of the most important tasks of modern physics. Such searches can go directly by aiming at higher energies and looking for new particles beyond the SM, or indirectly by scrutinizing possible NP effects in high-luminosity measurements. While the direct searches have not observed any NP signals so far, the second approach has provided some interesting hints of NP in several decay channels of the beauty mesons. One of the most exciting hints is the persistent excess of the measured branching fractions of the semileptonic decays $B\to D^{(*)}\tau\bar\nu_\tau$ over the SM prediction, which may imply violation of lepton flavor universality (LFU), and is widely known in the literature as ``the $R_{D^{(\ast)}}$ puzzle"~\cite{Ciezarek:2017yzh}. 

The ratios of branching fractions $R_{D^{(\ast)}} \equiv 
\mathcal{B}(\bar{B}^0 \to D^{(\ast)} \tau^- \bar{\nu}_{\tau})/\mathcal{B}(\bar{B}^0 \to D^{(*)} \ell^- \bar{\nu}_{\ell})$, where $\ell = e,\,\mu$, are often considered in order to reduce the hadronic uncertainties and to cancel the dependence on the Cabibbo--Kobayashi--Maskawa (CKM) matrix element $|V_{cb}|$. Independent measurements of $R_{D^{(\ast)}}$ by the {\it BABAR}~\cite{Lees:2012xj, Lees:2013uzd}, Belle~\cite{Huschle:2015rga, Sato:2016svk, Hirose:2016wfn}, and LHCb~\cite{Aaij:2015yra, Aaij:2017uff} collaborations showed a combined excess of about $4\sigma$ over the SM prediction, based on the analysis~\cite{Amhis:2016xyh} of the Heavy Flavor Averaging Group (HFLAV) in summer 2018. Very recently, the Belle collaboration reported a new measurement of the ratios of $R_{D^{(\ast)}}$~\cite{Abdesselam:2019dgh}. Their results (first presented at Moriond~2019)
\be
R_D=0.307\pm 0.037\,({\rm stat})\pm 0.016\,({\rm syst}),\qquad
R_{D^\ast}=0.283\pm 0.018\,({\rm stat})\pm 0.014\,({\rm syst}),
\en

\noindent
agree with the average SM predictions~\cite{Amhis:2016xyh, Bigi:2016mdz, Bernlochner:2017jka, Bigi:2017jbd, Jaiswal:2017rve} 
\be
R_D=0.299\pm 0.003,\qquad
R_{D^*}=0.258\pm 0.005,
\en
\noindent
within $0.2\sigma$ and $1.1\sigma$, respectively. The inclusion of these new results reduces the overall tension with the SM from $4\sigma$ to $3.1\sigma$, and the global average values now read~\cite{Amhis:2016xyh}

\be
R_D=0.340\pm 0.030,\qquad
R_{D^*}=0.295\pm 0.014.
\en

Even though the tension is now somehow reduced, the puzzle remains unsolved and attractive. One of the reasons is that similar anomalies also appear in other $B$ meson decays  (see, e.g.,~\cite{Bifani:2018zmi} for a recent review). In particular, the recent LHCb measurement~\cite{Aaij:2017tyk} of the ratio of branching fractions

\be
R_{J/\psi}\equiv \frac{\mathcal{B}(B_c \to J/\psi \tau\nu)}{\mathcal{B}(B_c \to J/\psi \mu\nu)}=0.71\pm 0.17\,(\rm stat)\pm 0.18\,(\rm syst)
\en

\noindent
also exceeds the SM predictions~\cite{Cohen:2018dgz, Leljak:2019eyw, Murphy:2018sqg} at about $1.5\sigma$. It~is important to note that the decays $B_c \to J/\psi \ell\nu$ and $B \to D^{(\ast)} \ell\nu$ are described by the same transition $b\to c\ell\nu$ at the quark level. 
The excess of $R_{J/\psi}$ over the SM predictions therefore implies hints of NP in the $b\to c\tau\nu_\tau$ transition, once again. It~also suggests the consideration of the decay $B_c \to \eta_c \tau\nu$ as a promising probe of NP.

The $R_{D^{(\ast)}}$ and $R_{J/\psi}$ puzzles have been the motivation of a huge number of theoretical studies, which can be divided into two basic categories: Specific models of NP and general effective Lagrangian approaches. The first approach explains the discrepancies by assuming the participation of additional mediators beyond the SM, such as charged Higgs bosons, $W^\prime$ boson, leptoquarks, etc., in the given process. Such models are well constructed, and the new mediators have some definite properties that can be tested by experiments. However, at the same time, they suffer from stringent experimental constraints coming from various processes, also including direct searches at the Large Hadron Collider. Details on these models can be found in the recent papers~\cite{Angelescu:2018tyl, Yan:2019hpm, Bansal:2018nwp, Li:2018rax, Marzo:2019ldg, Greljo:2018ogz, Gomez:2019xfw} and references therein. In the second approach, one starts with a general effective Hamiltonian for the weak $b\to c\ell\nu$ transition that includes both the SM and beyond-SM contributions in the form of dimension-six four-fermion operators. Experimental constraints on various physical observables in the decays are then used to discriminate between different NP scenarios. This approach is more general and exploratory in the sense that it may provide important insights for further construction of NP models if the discrepancy with the SM is confirmed. There is a large number of analyses using this approach in the literature. We therefore mention here only a few pioneering studies~\cite{Fajfer:2012vx, Datta:2012qk, Tanaka:2012nw, Biancofiore:2013ki, Dutta:2013qaa} as well as very recent papers~\cite{Murgui:2019czp, Kim:2018hlp, Hu:2018veh, Asadi:2019xrc, Blanke:2019qrx, Shi:2019gxi}.

Following the general Hamiltonian approach and using the hadronic form factors obtained in the covariant constituent quark model (CCQM), we have studied the $B$-meson anomalies in a series of papers~\cite{Ivanov:2016qtw, Ivanov:2017mrj, Tran:2018kuv, Gutsche:2018nks}. We have shown in detail how various polarization observables could help distinguish between NP contributions. In particular, we found that the longitudinal polarization fraction $F_L^{D^*}$ of the $D^*$ meson in the decay $B^0 \to D^{\ast} \tau\nu_\tau$ is very sensitive to the scalar and tensor four-fermion operators, and the effects are opposite: The scalar operator enhances, while the tensor one lowers the value of $F_L^{D^*}$~\cite{Ivanov:2016qtw} (also see~\cite{Alok:2016qyh, Huang:2018nnq,Iguro:2018vqb}). Recently, the Belle collaboration reported their first measurement of the fraction $F_L^{D^*}$~\cite{Abdesselam:2019wbt}, and the result was quite curious. For the electron mode, their measured value $F_L^{D^*}({B}^0 \to D^{\ast} e^+\nu_e)=0.56\pm 0.02$ agrees very well with our prediction of 0.54~\cite{Ivanov:2015tru}, while for the tau mode, their result $F_L^{D^*}({B}^0 \to D^{\ast} \tau^+\nu_\tau)=0.60\pm 0.08\,(\rm stat)\pm 0.04\,(\rm syst)$ lies at about $1.6\sigma$ above our prediction of 0.46~\cite{Ivanov:2015tru}. Based on our analysis~\cite{Ivanov:2017mrj}, one sees that this enhancement, if confirmed, is a clear evidence of the scalar operator. 

Moreover, the longitudinal polarization of the tau in $B \to D^{\ast} \tau\nu_\tau$ was also observed for the first time in a recent experiment at  Belle~\cite{Hirose:2016wfn}. Even though the result $P_L^\tau = -0.38 \pm 0.51\,(\text{stat})^{+0.21}_{-0.16}\, (\text{syst})$ still suffers from large uncertainties, this observation gives a clear message that more accurate data will soon be available at Belle~II. In light of the new experimental data, we redo the global fit for the NP Wilson coefficients with particular focus on the new measurement of $D^\ast$ polarization~\cite{Abdesselam:2019wbt} and its impact on the overall picture. The rest of the paper is organized as follows: In Section~\ref{sec:formalism}, we introduce some formalism concerning the semileptonic $B$ decay and the NP effective Hamiltonian. Section~\ref{sec:FF} is dedicated to the calculation of the form factors in the CCQM. Numerical results and their discussion are given in Section~\ref{sec:result}. Finally, we briefly conclude in Section~\ref{sec:sum}.

\section{Theoretical Framework}
\label{sec:formalism}

In the model-independent approach, the SM is extended by considering a general effective Hamiltonian for the quark-level transition $b \to c \ell \nu$ $(\ell=e,\mu,\tau)$ constructed from all dimension-six operators as follows $(i=L,R)$~\cite{Goldberger:1999yh,Buchmuller:1985jz,Grzadkowski:2010es}:
\bea
\label{eq:Heff}
\mathcal{H}_{eff} &=&\frac{4G_F V_{cb}}{\sqrt{2}} \Big(\mathcal{O}_{V_L}+
\sum\limits_{X=S_i,V_i,T_L} \delta_{\tau\ell}X\mathcal{O}_{X}\Big),
\ena
where the four-fermion operators $\mathcal{O}_{X}$ are given by
\bea
\mathcal{O}_{V_i} &=&
\left(\bar{c}\gamma^{\mu}P_ib\right)
\left(\bar{\ell}\gamma_{\mu}P_L\nu_{\ell}\right),\\
\mathcal{O}_{S_i} &=& \left(\bar{c}P_ib\right)\left(\bar{\ell}P_L\nu_{\ell}\right),
\\
\mathcal{O}_{T_L} &=& \left(\bar{c}\sigma^{\mu\nu}P_Lb\right)
\left(\bar{\ell}\sigma_{\mu\nu}P_L\nu_{\ell}\right).
\ena

Here, $\sigma_{\mu\nu}=i\left[\gamma_{\mu},\gamma_{\nu}\right]/2$, 
$P_{L,R}=(1\mp\gamma_5)/2$, and $X$s are the complex Wilson coefficients governing the NP contributions. The tensor operator with right-handed quark current simply does not contribute. One recovers the SM Hamiltonian by setting $V_{L,R}=S_{L,R}=T_L=0$. We have assumed that NP only couples to the third-generation leptons, and neutrinos are left-handed.

The matrix element of the semileptonic decays $B \to D^{(\ast)} \tau\nu_\tau$ and $B_c \to J/\psi(\eta_c) \tau\nu_\tau$ can be written in the following general form, where $P$ ($V$) denotes a pseudoscalar (vector) meson: 
\be
\mathcal{M}=\mathcal{M}_{\rm SM}+\sqrt{2} G_F V_{cb}\sum\limits_{X}
X\cdot
\langle V (P^\prime)
|\bar{c} \Gamma_X b
|P \rangle
\cdot
\bar\tau \Gamma_X \nu_\tau,
\en
where $\Gamma_X$ is the Dirac matrix corresponding to the operator $\mathcal{O}_X$. The hadronic part in the matrix element is parametrized by a set of invariant form factors depending on the momentum transfer squared $q^2$ between the two hadrons. For the $P\to P^\prime$ transition, one has
\bea
\langle P^\prime(p_2)
|\bar{c} \gamma^\mu b
|P(p_1) \rangle
&=& F_+(q^2) P^\mu + F_-(q^2) q^\mu,\\
\langle P^\prime(p_2)
|\bar{c}b
| P(p_1) \rangle &=& (m_1+m_2)F^S(q^2),\\
\langle P^\prime(p_2)|\bar{c}\sigma^{\mu\nu}(1-\gamma^5)b|P(p_1)\rangle 
&=&\frac{iF^T(q^2)}{m_1+m_2}\left(P^\mu q^\nu - P^\nu q^\mu 
+i \varepsilon^{\mu\nu Pq}\right),
\ena 
where $P=p_1+p_2$, $q=p_1-p_2$, $\varepsilon^{\mu\nu Pq}\equiv \varepsilon^{\mu\nu\alpha\beta}P_\alpha q_\beta$, and the mesons are on shell: $p_1^2=m_1^2=m_P^2$, and  $p_2^2=m_2^2=m_{V(P^\prime)}^2$. The $P\to V$ transition form factors are defined by
\bea
\langle V(p_2)
|\bar{c} \gamma^\mu(1\mp\gamma^5)b
|P(p_1) \rangle
&=& \frac{\epsilon^{\dagger}_{2\alpha}}{m_1+m_2}
\Big[ \mp g^{\mu\alpha}PqA_0(q^2) \pm P^{\mu}P^{\alpha}A_+(q^2)\nn
&&\pm q^{\mu}P^\alpha A_-(q^2) 
+ i\varepsilon^{\mu\alpha P q}V(q^2)\Big],\\
\langle V(p_2)
|\bar{c}\gamma^5 b
|P(p_1) \rangle &=& \epsilon^\dagger_{2\alpha}P^\alpha G^P(q^2),
\\
\langle V(p_2)|\bar{c}\sigma^{\mu\nu}(1-\gamma^5)b|P(p_1)\rangle
&=&-i\epsilon^\dagger_{2\alpha}\Big[
\left(P^\mu g^{\nu\alpha} - P^\nu g^{\mu\alpha} 
+i \varepsilon^{P\mu\nu\alpha}\right)G_1^T(q^2)\nn
&&+\left(q^\mu g^{\nu\alpha} - q^\nu g^{\mu\alpha}
+i \varepsilon^{q\mu\nu\alpha}\right)G_2^T(q^2)\nn
&&+\left(P^\mu q^\nu - P^\nu q^\mu 
+ i\varepsilon^{Pq\mu\nu}\right)P^\alpha\frac{G_0^T(q^2)}{(m_1+m_2)^2}
\Big],
\label{eq:ff}
\ena
where $\epsilon_2$ is the polarization vector
of the $V$ meson which satisfies the condition $\epsilon_2^\dagger\cdot p_2=0$.

The differential decay widths are written in terms of helicity amplitudes which, in turn, are combinations of the invariant form factors (see, e.g.,~\cite{Ivanov:2016qtw} for the full expressions). One has
\be
\frac{d\Gamma(P\to V(P^\prime)\tau\nu)}{dq^2}
=\frac{G_F^2|V_{cb}|^2|{\bf p_2}|q^2}{(2\pi)^3 12m_1^2}\Big(1-\frac{m^2_\tau}{q^2}\Big)^2\cdot {\cal H}_{tot}^{V(P^\prime)},
\label{eq:distr1}
\en
where
\bea
{\cal H}_{tot}^{P^\prime}
&=&
|1+g_V|^2\left[|H_0|^2+\delta_\tau(|H_0|^2+3|H_t|^2) \right]+\frac{3}{2}|g_S|^2 |H_P^S|^2\nn
&&+ 3\sqrt{2\delta_\tau} {\rm Re}g_S H_P^S H_t
+8|T_L|^2 ( 1+4\delta_\tau) |H_T|^2
+12\sqrt{2\delta_\tau} {\rm Re}T_L H_0 H_T,\\[1.0em]
{\cal H}_{tot}^{V}
&=&(|1+V_L|^2+|V_R|^2)\left[\sum\limits_{n=0,\pm}|H_{n}|^2+\delta_\tau \left(\sum\limits_{n=0,\pm}|H_{n}|^2+3|H_{t}|^2\right) \right]+\frac{3}{2}|g_P|^2|H^S_V|^2\nn
&&-2 {\rm Re}V_R\big[(1+\delta_\tau) (|H_{0}|^2+2H_{+}H_{-})+3\delta_\tau |H_{t}|^2 \big]
-3\sqrt{2\delta_\tau} {\rm Re}g_P H^S_V H_{t}\nn
&&+8|T_L|^2 (1+4\delta_\tau)\sum\limits_{n=0,\pm}|H_T^n|^2
-12\sqrt{2\delta_\tau} {\rm Re}T_L\sum\limits_{n=0,\pm} H_{n}H_T^n.
\ena

Here, $\delta_\ell\equiv m^2_\ell/2q^2$ is the helicity-flip factor, and $|{\bf p_2}|=\lambda^{1/2}(m_1^2,m_2^2,q^2)/2m_1$ is the momentum of
the daughter meson in the rest frame of the parent meson. For simplicity, we have introduced $g_V\equiv V_L+V_R$, $g_S\equiv S_L+S_R$, and $g_P\equiv S_L-S_R$.  Note that in this paper, we do not consider interference terms between different NP operators since we assume the dominance of only one NP operator
besides the SM contribution. 

\begin{figure}[htbp]
\centering
\includegraphics[scale=0.4]{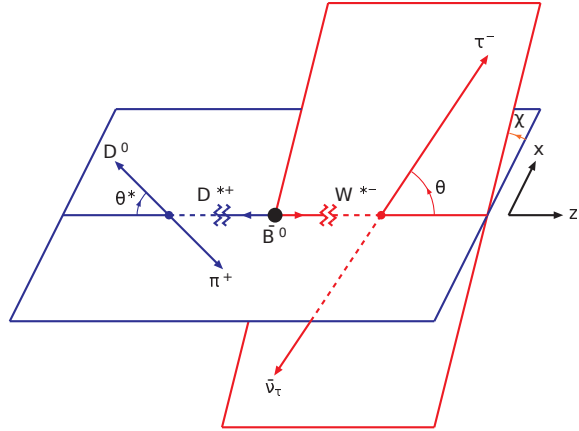}
\caption{Definition of the angles $\theta$, $\theta^\ast$, and $\chi$ in
the cascade decay $\bar B^0\to D^{\ast+}(\to D^0\pi^+)\tau^-\bar\nu_\tau$.}
\label{fig:bdangl}
\end{figure}

The polarization of the $D^\ast$ meson can be studied by considering the cascade decay ${B}^0\to D^{\ast}(\to D^0\pi)\ell\bar{\nu}_\ell$. The fourfold differential decay distribution is written in terms of the momentum transfer squared $q^2$, two polar angles, $\theta$ and $\theta^\ast$ in the dilepton and $D^*$ rest frames, respectively, and one azimuthal angle $\chi$, which are defined in Figure~\ref{fig:bdangl}. One has
\be
\frac{d^4\Gamma({B}^0\to D^{\ast}(\to D^0\pi)\ell\bar{\nu}_\ell)}
     {dq^2 d\cos\theta d\chi d\cos\theta^\ast} 
=\frac{9}{8\pi}|N|^2\mathcal{B}(D^\ast\to D^0\pi)J(\theta,\theta^\ast,\chi),
\label{eq:distr4}
\en
where
\be
|N|^2=
\frac{G_F^2 |V_{cb}|^2 |{\bf p_2}| q^2}{(2\pi)^3 12 m_1^2}\Big(1-\frac{m^2_\ell}{q^2}\Big)^2.
\en

The full angular distribution $J(\theta,\theta^\ast,\chi)$ is expanded on a trigonometric basis as follows:
\bea
\lefteqn{J(\theta,\theta^\ast,\chi)}\nn
&=& J_{1s}\sin^2\theta^\ast + J_{1c}\cos^2\theta^\ast
+(J_{2s}\sin^2\theta^\ast + J_{2c}\cos^2\theta^\ast)\cos2\theta\nn
&&+J_3\sin^2\theta^\ast \sin^2\theta \cos2\chi
+J_4\sin2\theta^\ast \sin2\theta \cos\chi\nn
&&+J_5\sin2\theta^\ast \sin\theta \cos\chi
+(J_{6s}\sin^2\theta^\ast+J_{6c}\cos^2\theta^\ast)\cos\theta\nn
&&+J_7\sin2\theta^\ast \sin\theta \sin\chi
+J_8\sin2\theta^\ast \sin2\theta \sin\chi
+J_9\sin^2\theta^\ast \sin^2\theta \sin2\chi ,
\label{eq:angular}
\ena
where $J_{i(a)}$ $(i=1,\dots,9; a=s,c)$ are angular coefficient functions, explicit expressions of which can be found in~\cite{Ivanov:2016qtw}. A novel model-independent method for measuring the angular coefficients was recently discussed in~\cite{Hill:2019zja}. In this paper, we are interested in the polarization of the $D^*$ meson, for which we need only $J_{1s(c)}$ and $J_{2s(c)}$. One has
%%%
\bea
4J_{1s} &=&
\frac{3+2\delta_\tau}{4} (|1+V_L|^2+|V_R|^2)(|H_{+}|^2+|H_{-}|^2)-(3+2\delta_\tau) {\rm Re}V_RH_{+}H_{-}\nn
&&-8\sqrt{2\delta_\tau} {\rm Re}T_L (H_{+}H_T^+ +H_{-}H_T^-)+4(1+6\delta_\tau) |T_L|^2 (|H_T^+|^2+|H_T^-|^2)
,\\
4J_{1c} &=&2|S_R-S_L|^2 |H^S_V|^2+4\sqrt{2\delta_\tau} {\rm Re} (S_R-S_L) H^S_V H_{0t}\nn
&&+(|1+V_L|^2+|V_R|^2-2{\rm Re}V_R)\Big[4\delta_\tau |H_{t}|^2+\Big(1+2\delta_\tau\Big)|H_{0}|^2\Big]\nn
&&-16\sqrt{2\delta_\tau} {\rm Re}T_L H_{0}H_T^0+16(1+2\delta_\tau)|T_L|^2|H_T^0|^2,\\
4J_{2s} &=&\frac{1}{4}(1-2\delta_\tau)\Big[
 (|1+V_L|^2+|V_R|^2)(|H_{+}|^2+|H_{-}|^2)\nn
&&-4{\rm Re}V_RH_{+}H_{-} -16|T_L|^2(|H_T^+|^2 + |H_T^-|^2)
\Big],\\
4J_{2c} &=&(1-2\delta_\tau)\Big[-
(|1+V_L|^2+|V_R|^2-2{\rm Re}V_R)|H_{0}|^2+16|T_L|^2|H_T^0|^2\Big].
\ena
%%%
Note again that we do not consider interference terms between different NP operators. 

After an integration of Eq.~(\ref{eq:distr4}) over all angles, one obtains the familiar differential decay~rate
\be
\frac{d\Gamma({B}^0\to D^{\ast} \ell\bar\nu_\ell)}{dq^2} =
|N|^2 J_{\rm tot},\qquad J_{\rm tot}=3J_{1c}+6J_{1s}-J_{2c}-2J_{2s}.
\en
For convenience, we define a normalized full angular distribution 
$\widetilde J(\theta^\ast,\theta,\chi)$ as follows:
\be
\widetilde J(\theta^\ast,\theta,\chi)=\frac{9}{8\pi}\frac{J(\theta^\ast,\theta,\chi)}
{J_{\rm tot}}.
\label{eq:normdis}
\en

One can easily check that the normalized angular distribution 
$\widetilde J(\theta^\ast,\theta,\chi)$ integrates to $1$ after
$\cos\theta^\ast,\,\cos\theta$, and $\chi$ integrations. By integrating 
Eq.~(\ref{eq:distr4}) over $\cos\theta$ and 
$\chi$, one obtains the hadron-side $\cos\theta^\ast$ distribution, whose normalized form can be written as
\be
\widetilde J(\theta^\ast)=\frac34\left(2F_L(q^2)\cos^2\theta^\ast+F_T(q^2)\sin^2\theta^\ast\right),
\en
where $F_L(q^2)$ and $F_T(q^2)$ are the polarization fractions of the $D^\ast$ meson and are defined by
\be
F_L(q^2)=\frac{3J_{1c}-J_{2c}}{J_{\rm tot}},\qquad F_T(q^2)=\frac{6J_{1s}-2J_{2s}}{J_{\rm tot}},\qquad F_L(q^2)+F_T(q^2)=1.
\en

%%%%%%%%%%%%%%%%%%%%%%%%%%%%%%%%%%%%%%%%%%
\section{Form Factors in the Covariant Constituent Quark Model}
\label{sec:FF}

The covariant constituent quark model has been developed by our group in a series of papers (see, e.g.,~\cite{Branz:2009cd, Ivanov:2011aa}). We only mention here some important features of the model for completeness. More detailed descriptions of the model and the calculation techniques can be found in~\cite{Branz:2009cd, Ivanov:2011aa, Dubnicka:2018gqg, Gutsche:2019wgu, Gutsche:2015mxa}.
In the CCQM, the interaction Lagrangian of a meson $M$ with its constituent quarks is constructed from the meson field $M(x)$ and
the interpolating quark current $J_{M}(x)$:
\bea
{\cal L}_{\rm int} &=& g_{M } M (x) J_{M}(x)+{\rm H.c.},\\
J_{M}(x)  &=&  \int\!\! dx_1\!\!\int\!\! dx_2\: 
F_{M}(x;x_1,x_2)\bar q_2^a(x_2)\Gamma_M q_1^a(x_1),
\ena
where $\Gamma_M=I$, $\Gamma_M=\gamma_5$, and $\Gamma_M=\gamma^\mu$ for a scalar, a pseudoscalar, and a vector meson, respectively. The quark--meson coupling $g_M$ is determined by using the compositeness condition $Z_M=0$, where  $Z_M$ is the wave function renormalization constant of the meson. 

Nonlocality of the quark--meson interaction is characterized by the vertex function $F_M(x;x_1,x_2)$, whose form reads
\be
F_M (x;x_1,x_2) =  
\delta^{(4)} \left(x-w_1 x_1-w_2 x_2 \right)\:
\Phi_M\left(\left(x_1-x_2\right)^2\right),
\en
where $w_i=m_{q_i}/(m_{q_i}+m_{q_j})$ and $(i,j=1,2)$, so that $w_{1}+w_{2}=1$. This form of the vertex function satisfies the translational invariance. It~has been shown in our previous work that the concrete form of the function $\Phi_M\left(\left(x_{1}-x_{2}\right)^{2}\right)$ has small effects on the final physical results. Therefore, for simplicity, it
is assumed to have the following Gaussian form in the momentum representation:
\be
\widetilde{\Phi}_{M}\left(-p^2\right) =
\exp\left( p^2/\Lambda_{M}^2\right).
\en
The parameter $\Lambda_{M}$ is a free parameter of the model that characterizes the finite size of the meson. 

In the framework of the CCQM, hadronic matrix elements are described by Feynman diagrams which are written as convolutions of quark propagators and vertex functions. Regarding the quark propagators $S_q$, we use the Fock--Schwinger representation as follows:
\be
S_q(k) = (m_{q}+\not\! k)
\int\limits_0^\infty\! d\alpha \exp[-\alpha(m_{q}^2-k^2)].
\en

The $B^0\to D^{(\ast)}$ and $B_c\to J/\psi(\eta_c)$ invariant form factors are calculated from the corresponding one-loop quark diagrams. More details regarding the one-loop evaluation techniques can be found in~\cite{Ivanov:2015tru, Ivanov:2015woa, Soni:2018adu, Ivanov:2019nqd}, where semileptonic meson decays were computed. A form factor $F$ can be finally written in the form of a threefold integral
\be
F   = \int\limits_0^{1/\lambda^2}\!\! dt\, t
\!\! \int\limits_0^1\!\! d\alpha_1
\!\! \int\limits_0^1\!\! d\alpha_2  \,
\delta\Big(1 -  \alpha_1-\alpha_2 \Big) 
f(t\alpha_1,t\alpha_2),
\label{eq:3fold}
\en
where $f(t\alpha_1,t\alpha_2)$ is the resulting integrand corresponding to the form factor $F$, and $\lambda$ is a universal infrared cutoff parameter that guarantees the absence of branching points
corresponding to the creation of free quarks. 

Before presenting the results for the form factors, it should be mentioned that the model contains several free parameters: The constituent quark masses, the hadron size parameters $\Lambda_H$, and the universal infrared cutoff parameter $\lambda$. These parameters are determined from a least-squares fit to available experimental data and some lattice calculations. Those parameters 
involved in this paper are given by (in Gigaelectron Volts (GeV))~\cite{Ivanov:2011aa} 
\be
\begin{tabular}{ c| c| c| c|  c| c|  c|c|c|c|c}
 $m_{u/d}$ &  $m_s$  &  $m_c$  & $m_b$ 
&  $\lambda$  &  $\Lambda_{B}$   & $\Lambda_{B_c}$ & $\Lambda_{D}$ & $\Lambda_{D^*}$ & $\Lambda_{J/\psi}$ & $\Lambda_{\eta_c}$
\\\hline
 0.241  &  0.428 &  1.67 &
 5.04  &0.181 &  1.96  & 2.73 & 1.60 & 1.53 & 1.74 & 3.78
\end{tabular}
.
\label{eq:modelparam}
\en

Once the free parameters are fixed, the CCQM can be used as a strong tool to calculate hadronic quantities. The model has been successfully applied for numerous studies of not only mesons, but also baryons and other multiquark states.

In the CCQM, the form factors  are calculable in the full kinematical momentum 
transfer region $0\le q^2 \le q^2_{max}=(m_1-m_2)^2$. We use \textsc{fortran} codes from the Numerical Algorithms Group (NAG)
 library to do the numerical calculation of the threefold integrals in Eq.~(\ref{eq:3fold}). The calculated results are then interpolated by a double-pole parametrization
\be
F(q^2)=\frac{F(0)}{1 - a s + b s^2}, \quad s=\frac{q^2}{m_1^2}. 
\label{eq:ff-para}
\en
The parameters of the form factors for the $B^0\to D^{(\ast)}$ and $B_c\to J/\psi(\eta_c)$ transitions are listed in Tables~\ref{tab:ffBD} and~\ref{tab:ffBcJpsi}, respectively. Zero-recoil (or $q^2_{\rm max}$) values of the form factors are also listed for further comparison. 

\begin{table}[htbp]
\caption{Parameters of the dipole approximation in Equation~(\ref{eq:ff-para}) for  $B^0 \to D^{(\ast)}$ form factors. Zero-recoil values of the form factors are also listed.}
\centering
\begin{tabular}{c|ccccccccccccc}
\hline\hline
\toprule
 & {\boldmath $A_0 $} & {\boldmath $A_+$} & {\boldmath $ A_-$} & {\boldmath $V$} 
 & {\boldmath $G^P$} & {\boldmath $G_0^T$} & {\boldmath $G_1^T$} & {\boldmath $G_2^T$} & {} & {\boldmath $F_+$} & {\boldmath $F_-$} & {\boldmath $F^S$} & {\boldmath $F^T$}
 \\
 \midrule
 \hline
{\boldmath $F(0)$} &  1.62 & 0.67  & $-0.77$ & 0.77 & $-0.50$ & $-0.073$ & 0.73 & $-0.37$ & {} &  0.79   & $-0.36$ &  0.80 & 0.77  
\\
{\boldmath $a$}    &  0.34 & 0.87  &  0.89 & 0.90 & 0.87 & 1.23 & 0.90 & 0.88 & {} &  0.75   &  0.77 &  0.22 & 0.76  
\\
{\boldmath $b$}    & $-0.16$ & 0.06 & 0.07 & 0.08 & 0.06 & 0.33 & 0.07 & 0.07 & {} &  0.04  & 0.05 & $-0.10$ & 0.04 
\\ 
{\boldmath $F(q^2_{\rm max})$} &  1.91 & 0.99  & $-1.15$ & 1.15 & $-0.74$ & $-0.13$ & 1.10 & $-0.55$ & {} &  1.14   & $-0.53$ &  0.89 & 1.11  \\
\bottomrule
\hline\hline
\end{tabular}
\label{tab:ffBD}
\end{table}
\unskip
\begin{table}[htbp]
\caption{Parameters of the dipole approximation in Equation~(\ref{eq:ff-para}) for  $B_c \to J/\psi(\eta_c)$ form factors. Zero-recoil values of the form factors are also listed.}
\centering
\begin{tabular}{c|ccccccccccccc}
\toprule
\hline\hline
 & {\boldmath $A_0 $} & {\boldmath $A_+$} & {\boldmath $ A_-$} & {\boldmath $V$} 
 & {\boldmath $G^P$} & {\boldmath $G_0^T$} & {\boldmath $G_1^T$} & {\boldmath $G_2^T$} & {} & {\boldmath $F_+$} & {\boldmath $F_-$} & {\boldmath $F^S$} & {\boldmath $F^T$}
 \\
\midrule
\hline
{\boldmath $F(0)$} &  1.65 & 0.55  & $-0.87$ & 0.78 & $-0.61$ & $-0.21$ & 0.56 & $-0.27$ & {} &  0.75   & $-0.40$ &  0.69 & 0.93  
\\
{\boldmath $a$}    &  1.19 & 1.68  &  1.85 & 1.82 & 1.84 & 2.16 & 1.86 & 1.91 & {} &  1.31   &  1.25 &  0.68 & 1.30  
\\
{\boldmath $b$}    & 0.17 & 0.70 & 0.91 & 0.86 & 0.91 & 1.33 & 0.93 & 1.00 & {} &  0.33  & 0.25 & $-0.12$ & 0.31 
\\ 
{\boldmath $F(q^2_{\rm max})$} &  2.34 & 0.89  & $-1.49$ & 1.33 & $-1.03$ & $-0.39$ & 0.96 & $-0.47$ & {} &  1.12   & $-0.59$ &  0.86 & 1.40  
\\
\bottomrule
\hline\hline
\end{tabular}
\label{tab:ffBcJpsi}
\end{table}

\section{Numerical Analysis}
\label{sec:result} 
%%%%%%%%%%%%%%%%%%%%%%%%%%%%%%%%%%%%%%%%%%
In this paper, we assume that, besides the SM contribution, only one NP operator appears at a time. We also assume that all NP Wilson coefficients appearing in Eq.~(\ref{eq:Heff}) are complex. The allowed regions for each of these coefficients are obtained by using experimental results for the ratios of branching fractions $R_D=0.340\pm 0.030$, $R_{D^*}=0.295\pm 0.014$~\cite{Amhis:2016xyh}, $R_{J/\psi}=0.71\pm 0.25$~\cite{Aaij:2017tyk}, the upper limit ${\cal B}(B_c\to \tau\nu) \leq 10\,\%$ from the LEP1 data~\cite{Akeroyd:2017mhr}, and the longitudinal polarization fraction of the $D^*$ meson $F_L^{D^*}({B} \to D^{\ast} \tau\nu_\tau)=0.60\pm 0.09$~\cite{Abdesselam:2019wbt}. Within the SM, our quark model predicts $R_D=0.267$, $R_{D^*}=0.238$, $R_{J/\psi}=0.243$, ${\cal B}(B_c\to \tau\nu) = 2.74\%$, and $F_L^{D^*}({B} \to D^{\ast} \tau\nu_\tau)=0.45$. The theoretical errors for our predictions are estimated to be of order of 10\%.

When the general effective Hamiltonian~(\ref{eq:Heff}) for the $b\to c\ell\nu_\ell$ transition is invoked, the pure leptonic decay of the $B_c$ meson with a tau in the final state is also affected. To be more specific, all NP operators except for the tensor one contribute to this channel. The  tauonic branching fraction of $B_c$ in the presence of NP is given by~\cite{Ivanov:2017hun}
\be
 \mathcal{B}(B_c \to \tau \nu)=
\frac{G_F^2}{8\pi}|V_{cb}|^2\tau_{B_c}m_{B_c}m_{\tau}^2\left(1-\frac{m_{\tau}^2}{m_{B_c}^2}\right)^2f_{B_c}^2
 \times
\left|
1-g_A+\frac{m_{B_c}}{m_\tau} \frac{f_{B_c}^P}{f_{B_c}}g_P
\right|^2,
\en
where $g_A\equiv V_R-V_L$, $g_P\equiv S_R-S_L$, $\tau_{B_c}$ is the $B_c$ lifetime, $f_{B_c}$ is the leptonic decay constant of $B_c$, and $f_{B_c}^P$ is a new  constant corresponding to the new quark current structure $\langle 0|\bar{q}\gamma_5b|B_c(p)\rangle = m_{B_c} f_{B_c}^P$. In~the CCQM, one obtains $f_{B_c}=489.3$~MeV and $ f_{B_c}^P=645.4$~MeV.

\begin{figure}[htbp]
\centering
\begin{tabular}{cc}
\includegraphics[scale=0.5]{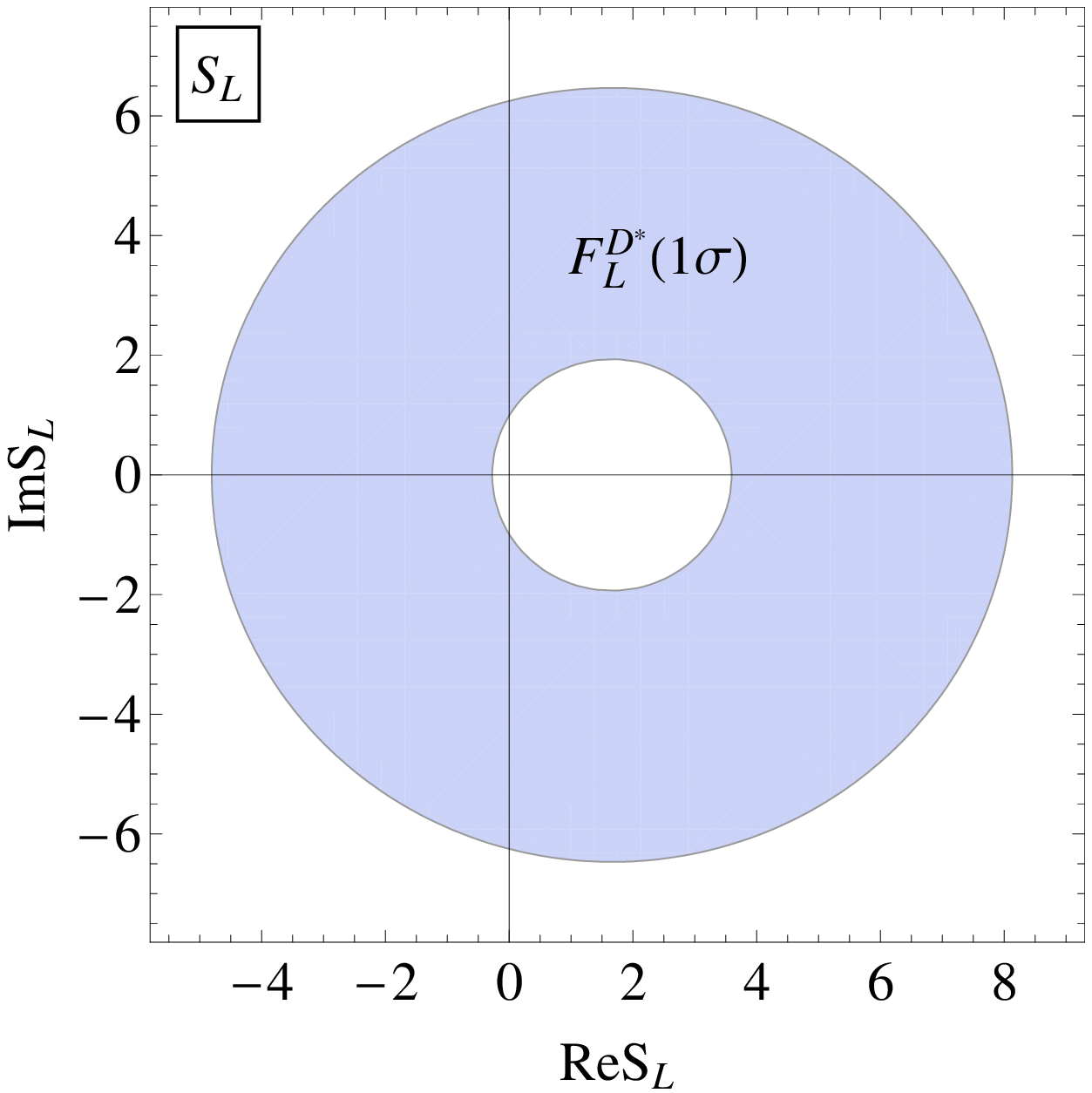}&
\includegraphics[scale=0.5]{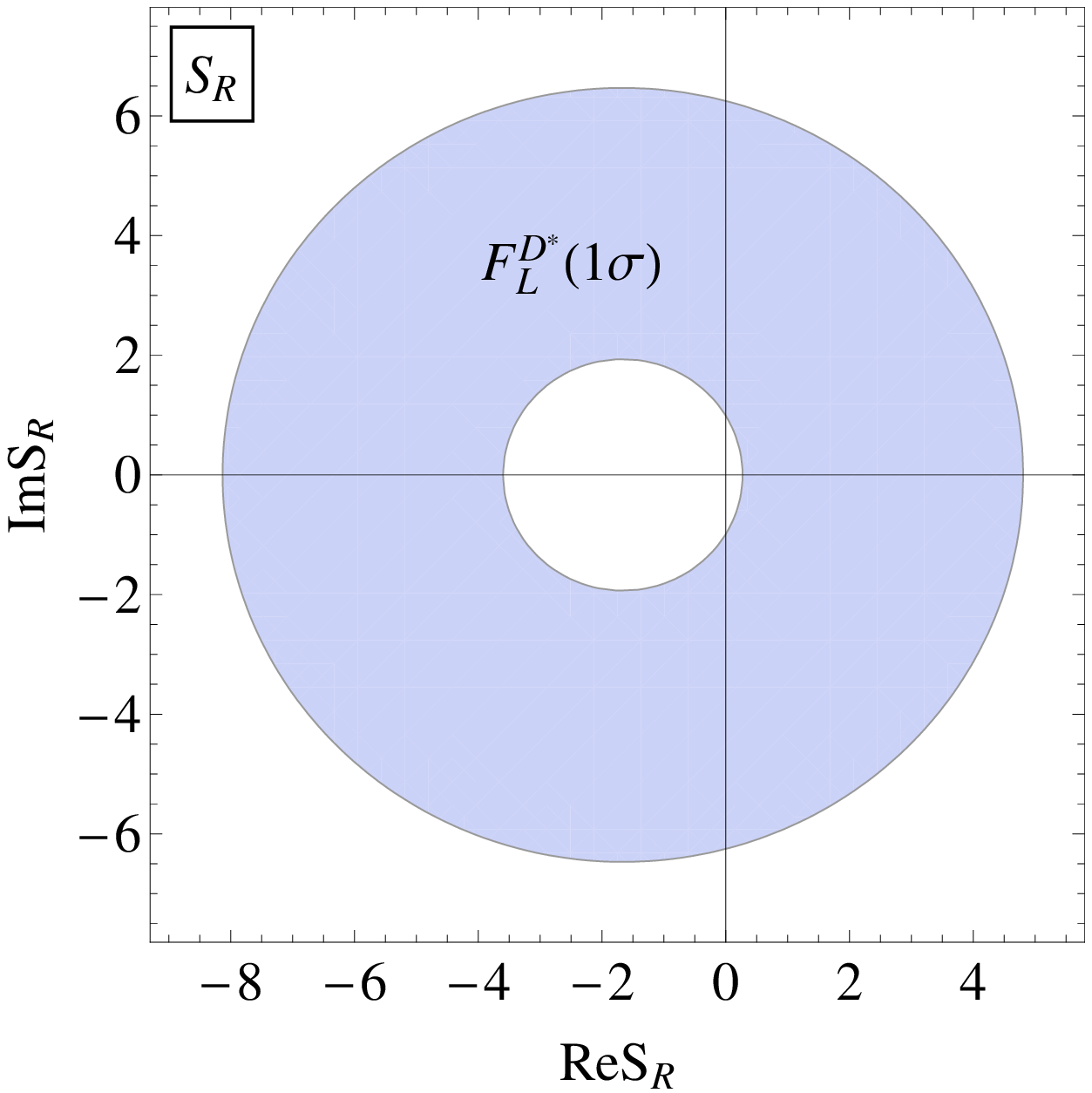}\\
\includegraphics[scale=0.5]{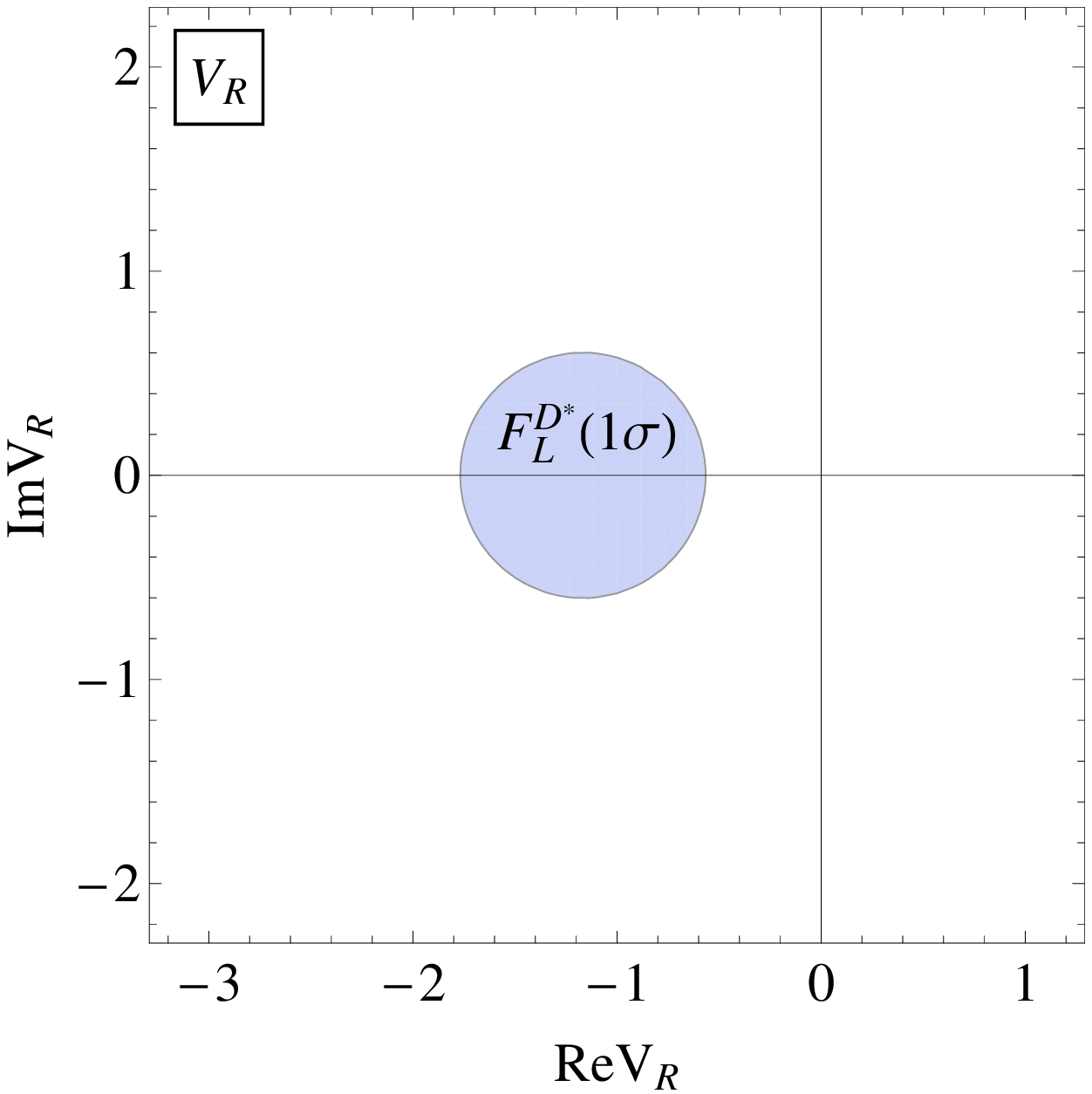}&
\includegraphics[scale=0.5]{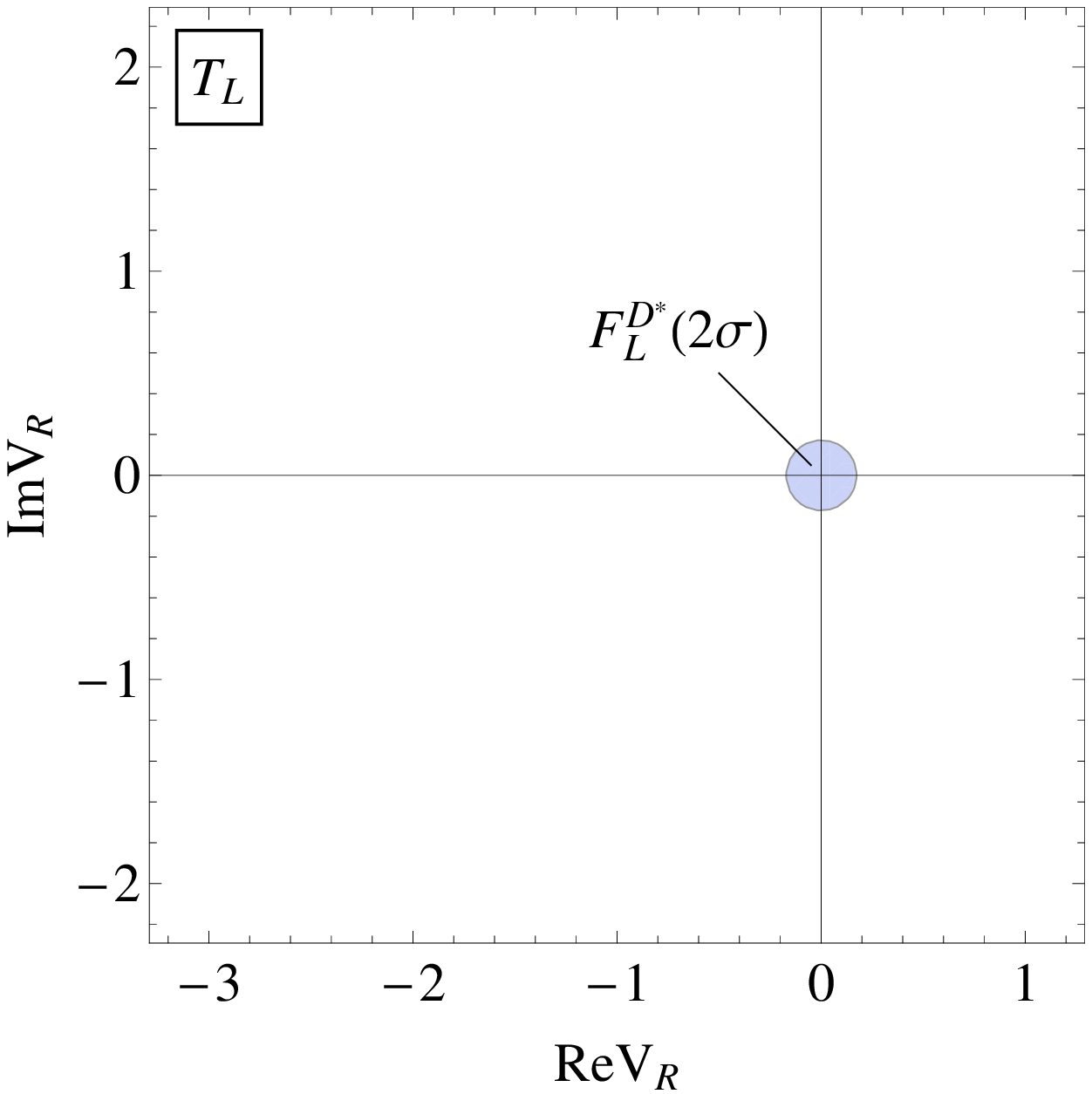}
\end{tabular}
\caption{Constraints on the complex Wilson coefficients $S_L$ (\textbf{upper, left}), $S_R$ (\textbf{upper, right}), $V_R$ (\textbf{lower, left}), and $T_L$ (\textbf{lower, right}) from the measurement of $F_L^{D^*}({B} \to D^{\ast} \tau\nu_\tau)$ within $1\sigma$ (for $S_L$, $S_R$, $V_R$) and $2\sigma$ (for $T_L$). The allowed regions are indicated in gray color.}
\label{fig:FLregion}
\end{figure}

First of all, we consider separately the new constraint coming from the recently measured longitudinal polarization $F_L^{D^*}({B} \to D^{\ast} \tau\nu_\tau)$ on the NP Wilson coefficients. The allowed regions of the last are shown in Figure~\ref{fig:FLregion}. It~is seen that the current data on $F_L^{D^*}$ prefer the scalar scenarios $S_{L,R}$, and the right-handed vector coefficient $V_R$ is still viable within $1\sigma$. Moreover, the most fruitful implication of the $F_L^{D^*}$ measurement is that it singly rules out the tensor operator at $1\sigma$, and also severely constrains it at $2\sigma$. This makes the explanation of the $b\to c\tau\nu$ anomalies based solely on the tensor interaction become less likely. 

\begin{figure}[htbp]
\centering
\begin{tabular}{cc}
\includegraphics[scale=0.5]{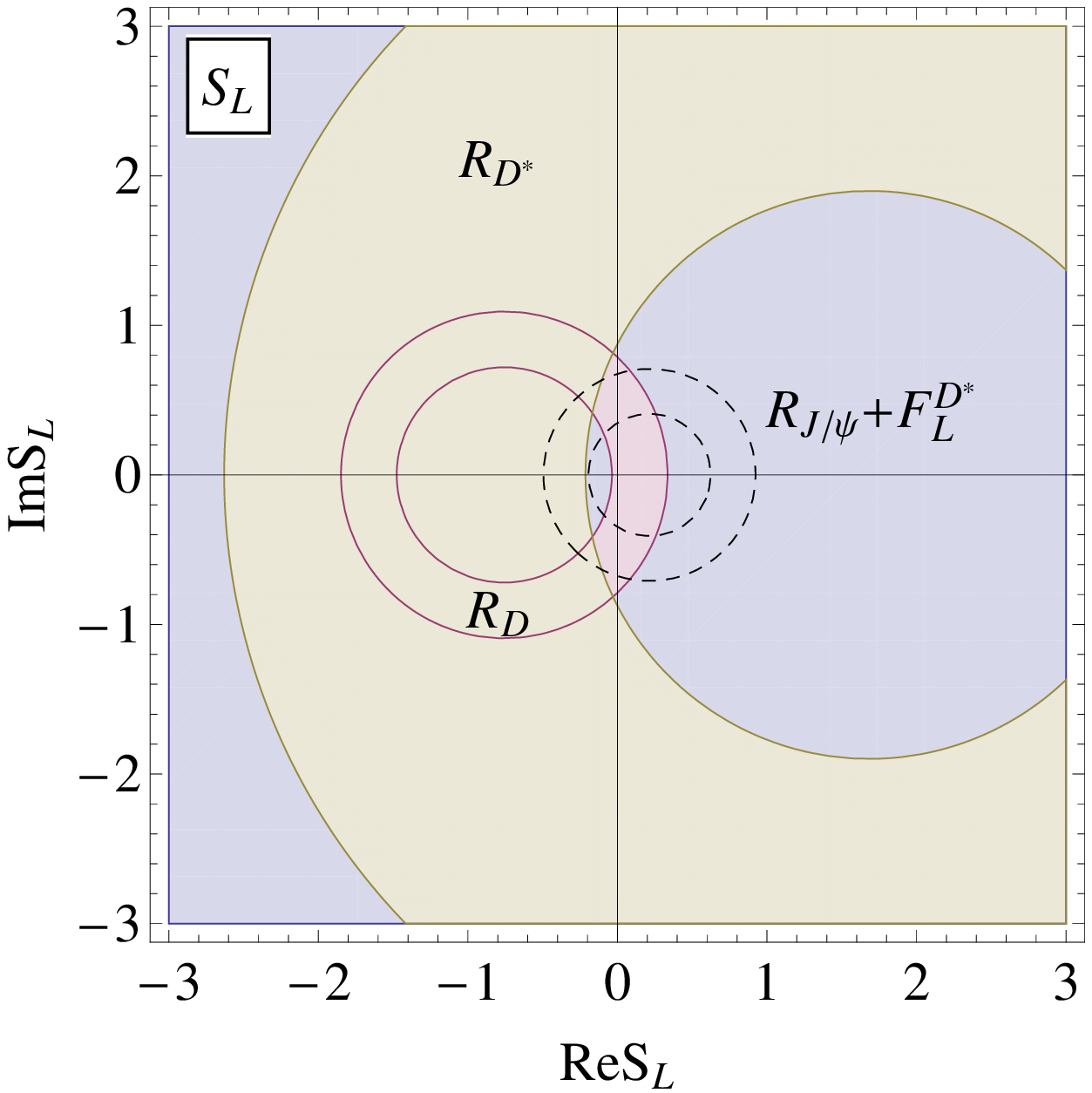}&
\includegraphics[scale=0.5]{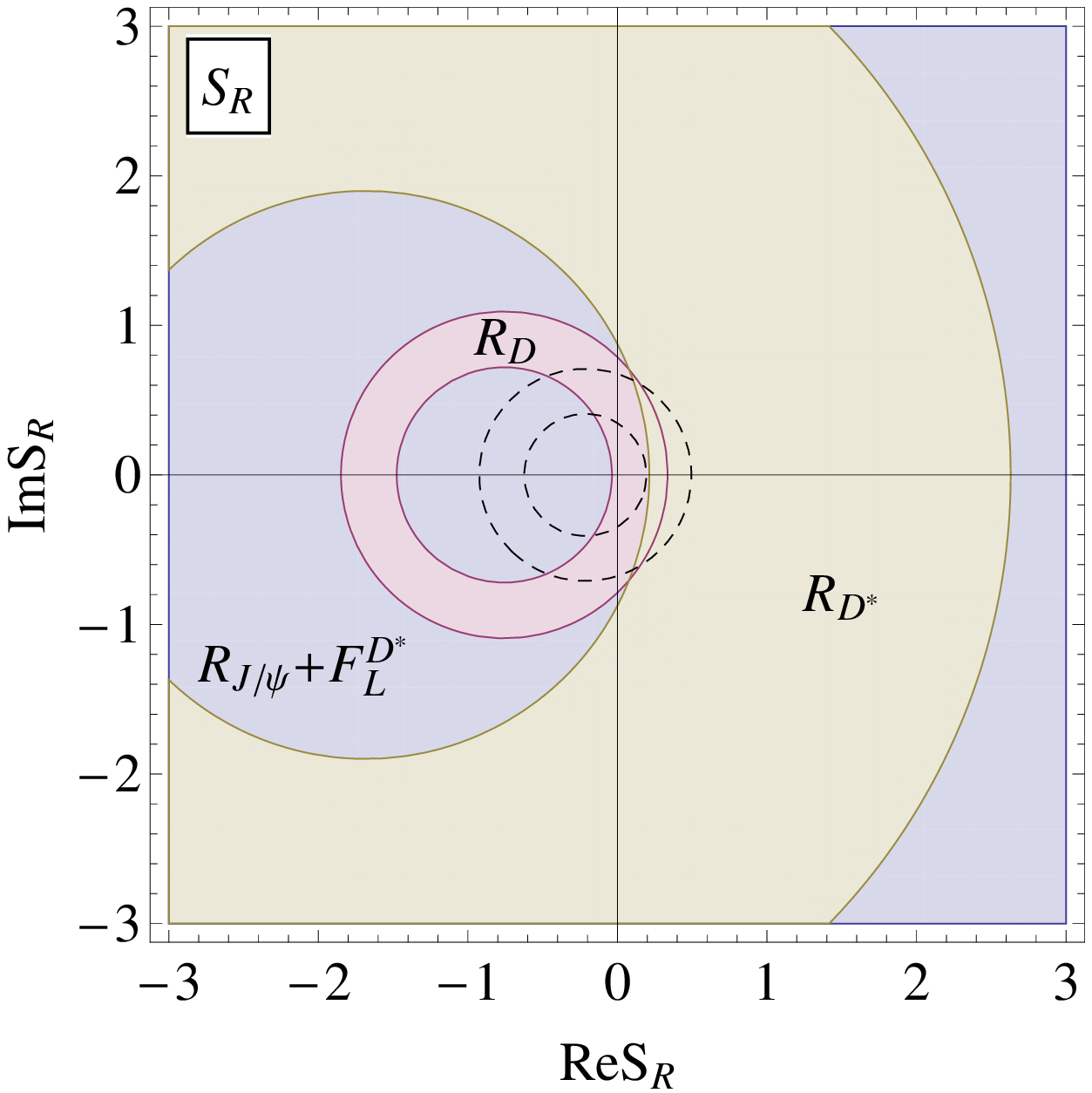}
\end{tabular}
\caption{Constraints on the complex Wilson coefficients $S_L$ (\textbf{left}) and $S_R$ (\textbf{right}) from the measurements of $R_D$, $R_{D^*}$, $R_{J/\psi}$, and $F_L^{D^*}({B} \to D^{\ast} \tau\nu_\tau)$ within $2\sigma$, and from the branching fraction ${\cal B}(B_c\to\tau\nu_\tau)$. The small (large) dash curve represents the constraint ${\cal B}(B_c\to\tau\nu_\tau)\leq 10\%$ (${\cal B}(B_c\to\tau\nu_\tau)\leq 30\%$).}
\label{fig:SLSR2sigma}
\end{figure}

In Figure~\ref{fig:SLSR2sigma}, we include all of the available constraints on the scalar scenarios at the level of $2\sigma$. The quick observation is that both $S_L$ and $S_R$ are excluded at $2\sigma$. However, more detailed notations should be made. Firstly, in both cases, the current data on $R_{J/\psi}$ and $F_L^{D^*}$ do not provide any additional effective constraints to those already given by $R_D$, $R_{D^*}$, and ${\cal B}(B_c\to\tau\nu_\tau)$. Secondly, the combination of the two well-measured ratios $R_D$ and $R_{D^*}$ prefers $S_L$ to $S_R$: $S_L$ is well allowed within $1\sigma$, while $S_R$ is  almost excluded at $2\sigma$. It~is seen that the branching fraction ${\cal B}(B_c\to\tau\nu_\tau)$ offers a very stringent constraint on possible scalar contributions in the $b\to c$ semitauonic transition: $S_L$ and $S_R$ are ruled out mainly by the constraint ${\cal B}(B_c\to\tau\nu_\tau)\leq 10\%$. If one uses the more relaxed constraint ${\cal B}(B_c\to\tau\nu_\tau)\leq 30\%$ obtained from the $B_c$ lifetime~\cite{Alonso:2016oyd}, then one finds that $S_L$ and $S_R$ are still available at $2\sigma$, but only to a small extent. Better experimental data for ${\cal B}(B_c\to\tau\nu_\tau)$ are therefore highly expected.

\unskip
%%%%
\begin{figure}[htbp]
\centering
\begin{tabular}{ccc}
\hspace{-1cm}
\includegraphics[scale=0.4]{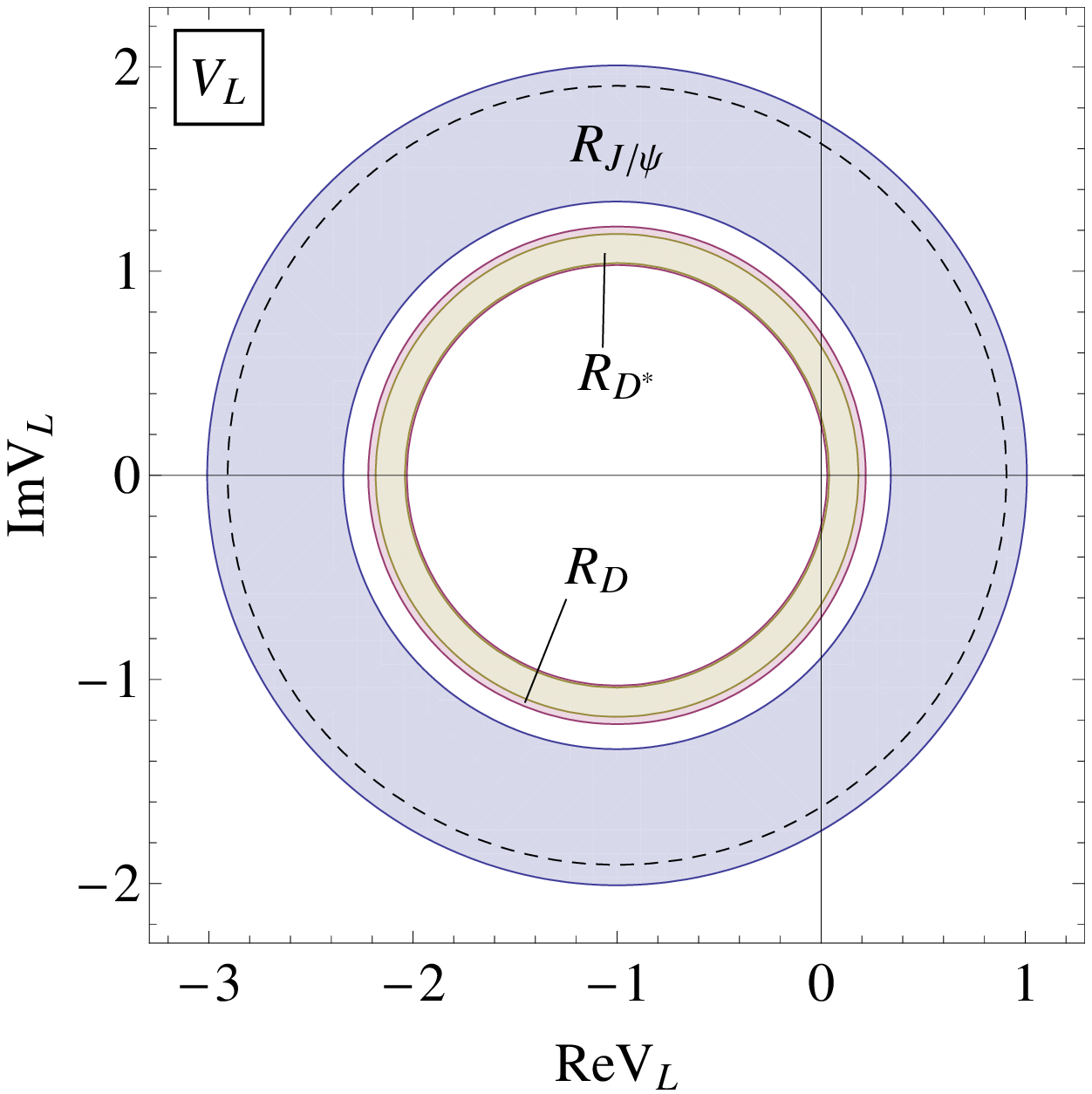}&
\includegraphics[scale=0.4]{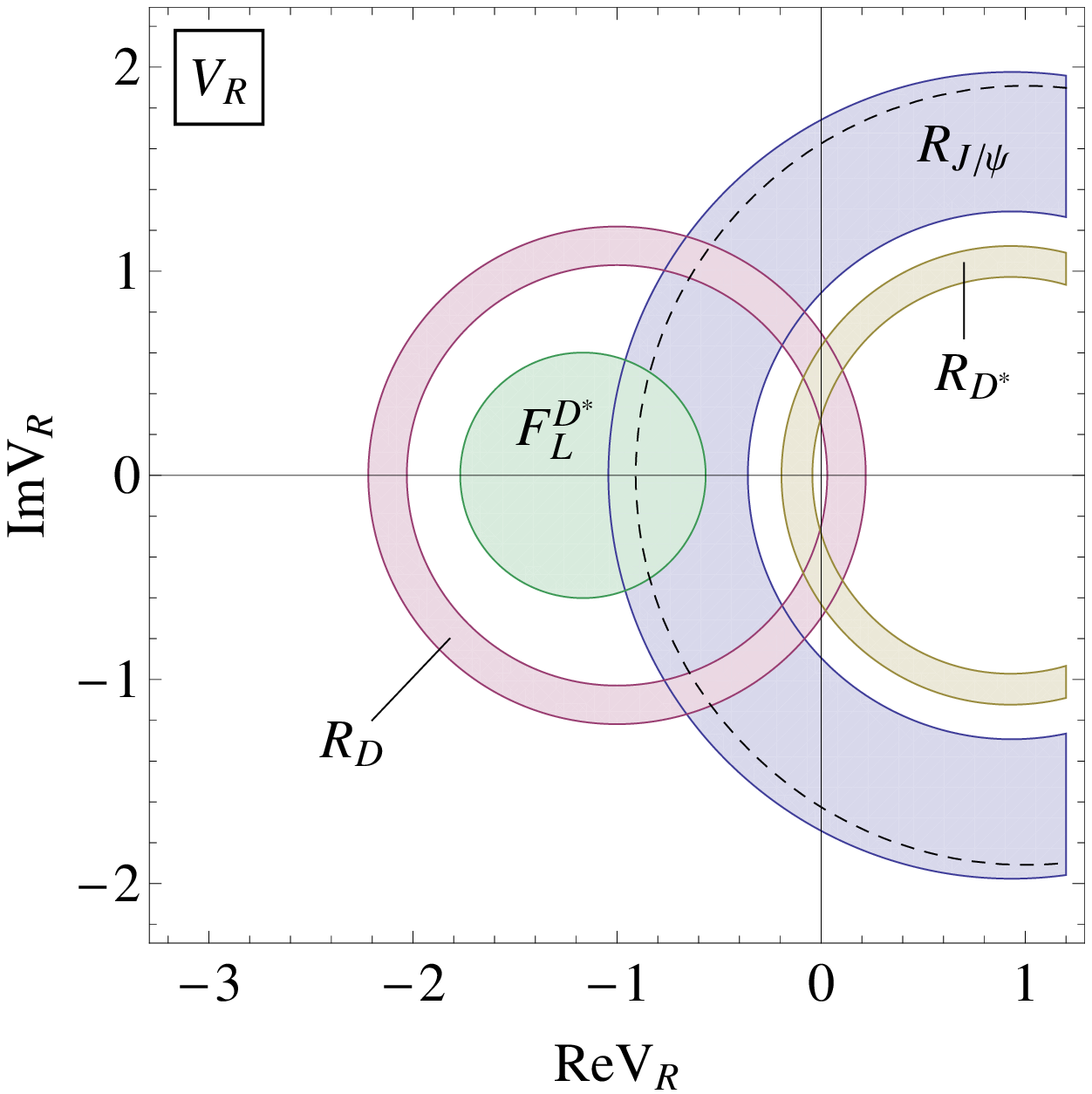}&
\includegraphics[scale=0.4]{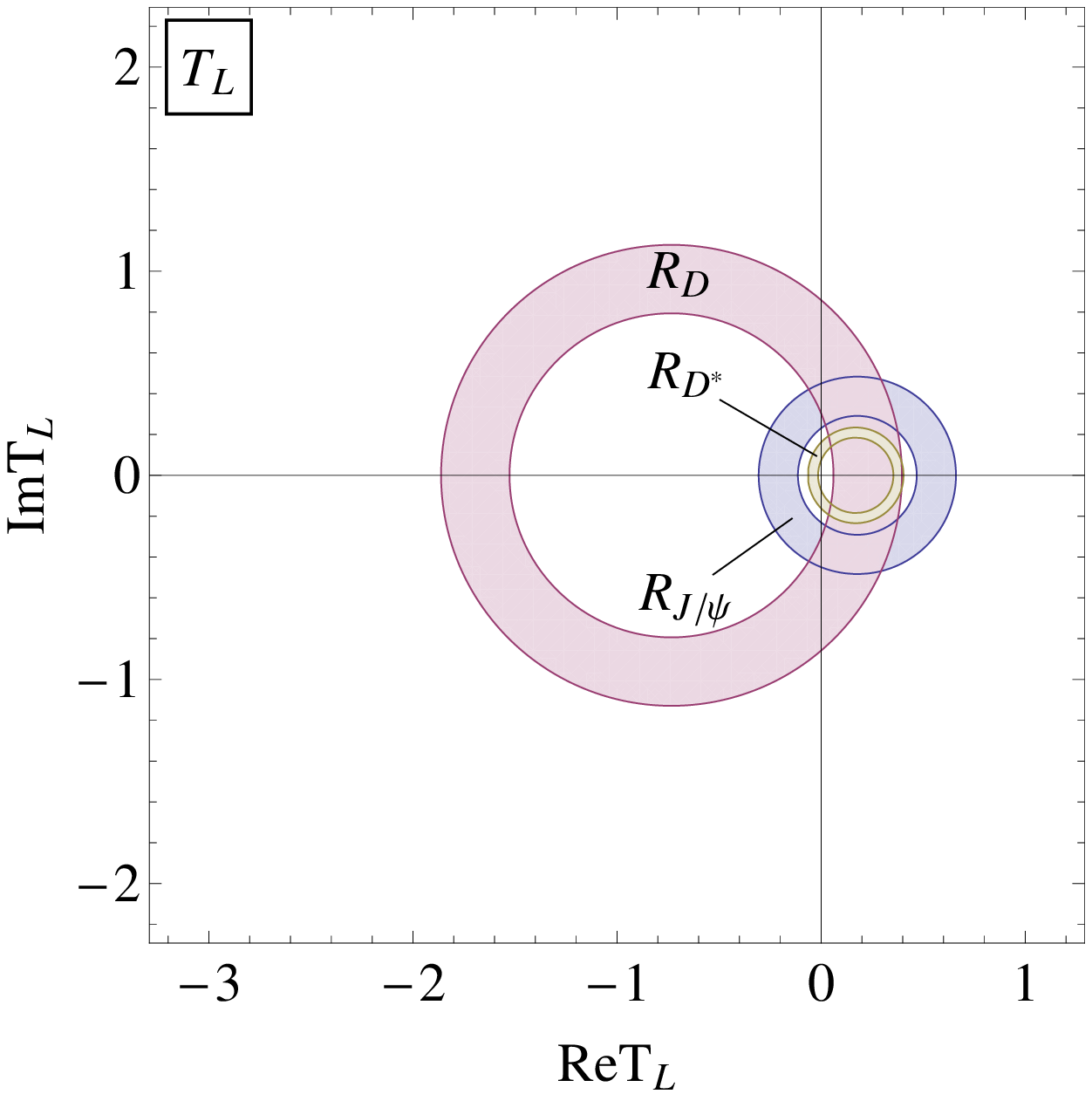}
\end{tabular}
\caption{Constraints on the complex Wilson coefficients $V_L$ (\textbf{left}), $V_R$ (\textbf{center}), and $T_L$ (\textbf{right}) from the measurements of $R_D$, $R_{D^*}$, $R_{J/\psi}$, and $F_L^{D^*}({B} \to D^{\ast} \tau\nu_\tau)$ within $1\sigma$, and from the branching fraction ${\cal B}(B_c\to\tau\nu_\tau)\leq 10\%$ (dashed curve).}
\label{fig:VT1sigma}
\end{figure}

The constraints on the vector and tensor Wilson coefficients within $1\sigma$ are presented in Figure~\ref{fig:VT1sigma}. None of the three operators is allowed at $1\sigma$. Unlike the scalar scenarios, the constraint from the branching fraction ${\cal B}(B_c\to\tau\nu_\tau)$ on the vector operators is still far less strict than that from the ratios $R_D$, $R_{D^*}$, and $R_{J/\psi}$. For the tensor scenario, ${\cal B}(B_c\to\tau\nu_\tau)$ simply has no effect. At $1\sigma$, the tensor operator is ruled out either by the combined data for $R_D$, $R_{D^*}$, and $R_{J/\psi}$, or by the polarization fraction $F_L^{D^*}({B} \to D^{\ast} \tau\nu_\tau)$ alone (see Figure~\ref{fig:FLregion}). The new constraint from $F_L^{D^*}$ reduces the likelihood of the $V_R$ scenario, which is now disfavored by either the combination of $R_D$, $R_{D^*}$, and $R_{J/\psi}$, or the combination of $R_D$, $R_{D^*}$, and $F_L^{D^*}$.  Note that $F_L^{D^*}$ is independent of $V_L$, and therefore, the new measurement of $F_L^{D^*}$ does not change the situation regarding $V_L$.

\begin{figure}[htbp]
\centering
\begin{tabular}{ccc}
\hspace{-1cm}
\includegraphics[scale=0.4]{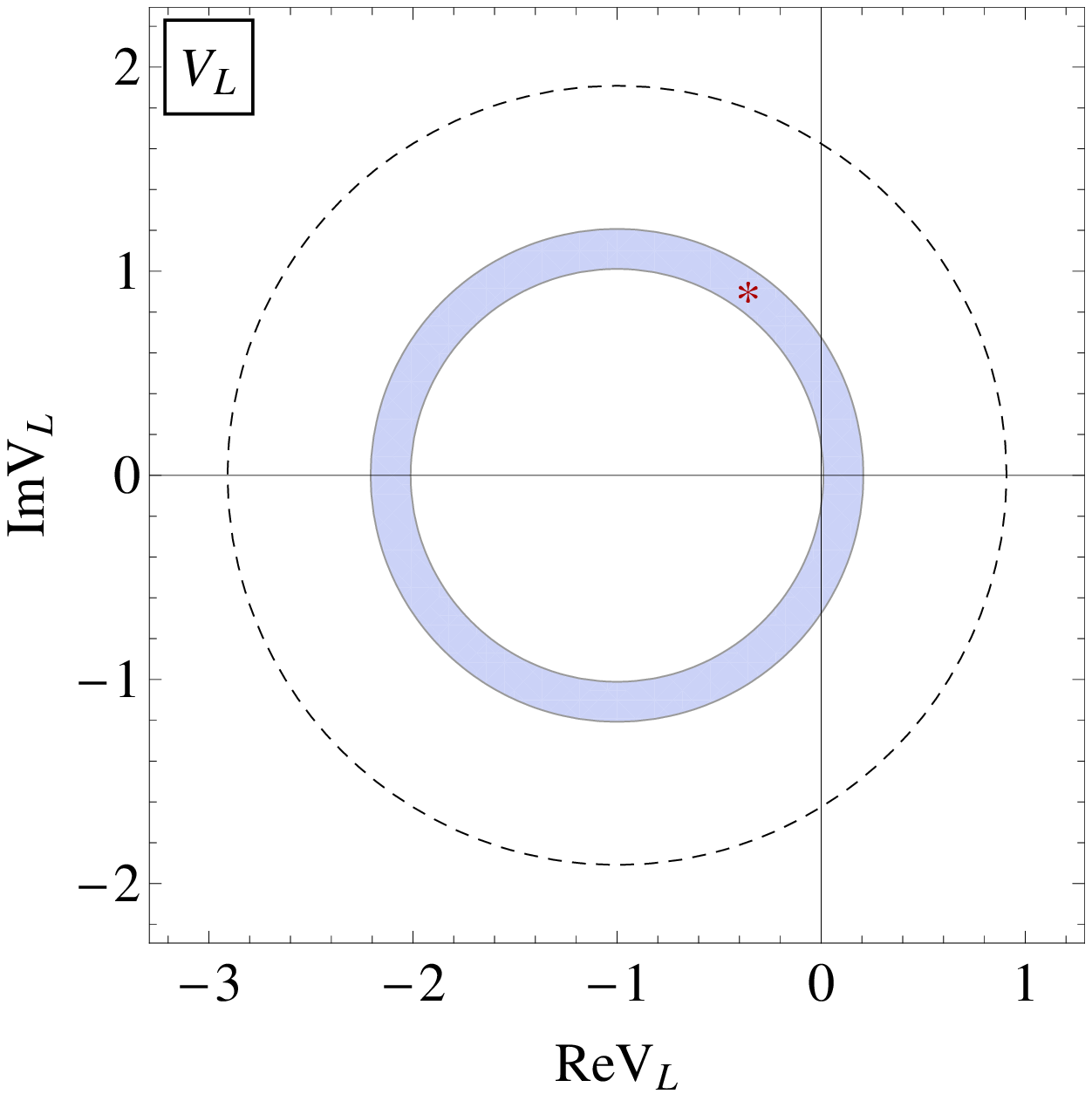}&
\includegraphics[scale=0.4]{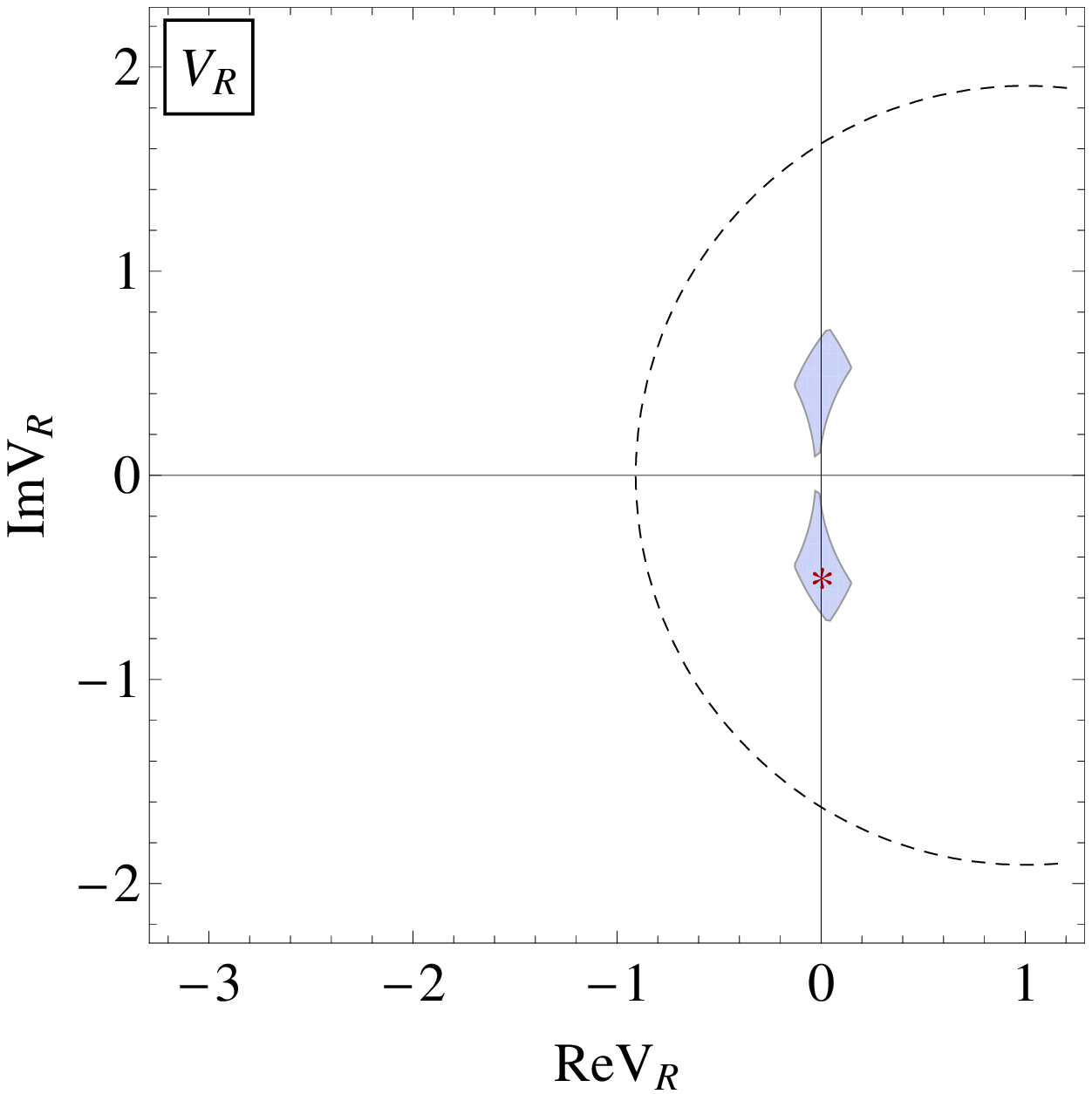}&
\includegraphics[scale=0.4]{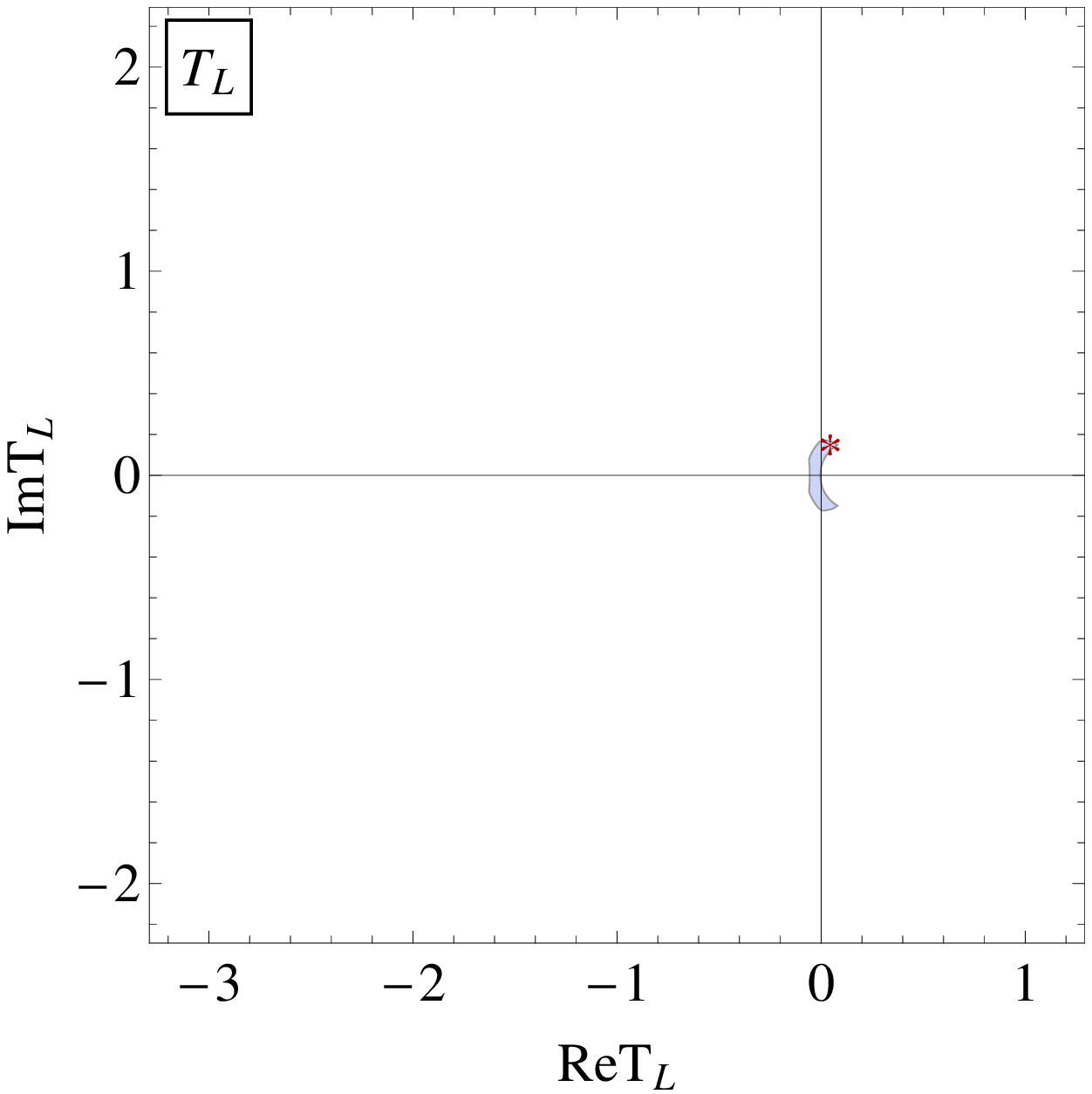}
\end{tabular}
\caption{Constraints on the complex Wilson coefficients $V_L$ (\textbf{left}), $V_R$ (\textbf{center}), and $T_L$ (\textbf{right}) from the measurements of $R_D$, $R_{D^*}$, $R_{J/\psi}$, and $F_L^{D^*}({B} \to D^{\ast} \tau\nu_\tau)$ within $2\sigma$, and from the branching fraction ${\cal B}(B_c\to\tau\nu_\tau)\leq 10\%$ (dashed curve). The allowed regions are indicated in gray color. The asterisk symbols indicate the best-fit values.}
\label{fig:VT2sigma}
\end{figure}

Finally, in Figure~\ref{fig:VT2sigma}, we show the allowed regions for $V_L$, $V_R$, and $T_L$ within $2\sigma$. In each region, we find a best-fit value and mark it with an asterisk. The best-fit values read
\be
V_L=-0.36+i\,0.92,\qquad
V_R=0.01-i\,0.48,\qquad
T_L=0.04+i\,0.17.
\label{eq:bestfit}
\en

The $2\sigma$ allowed regions together with these best-fit values can be used to analyze the effects of NP operators on various physical observables, as has been done in numerous papers, including our detailed analyses~\cite{Ivanov:2016qtw, Ivanov:2017mrj,Tran:2018kuv}. 

\begin{figure}[htbp]
\centering
\begin{tabular}{cc}
\includegraphics[scale=0.5]{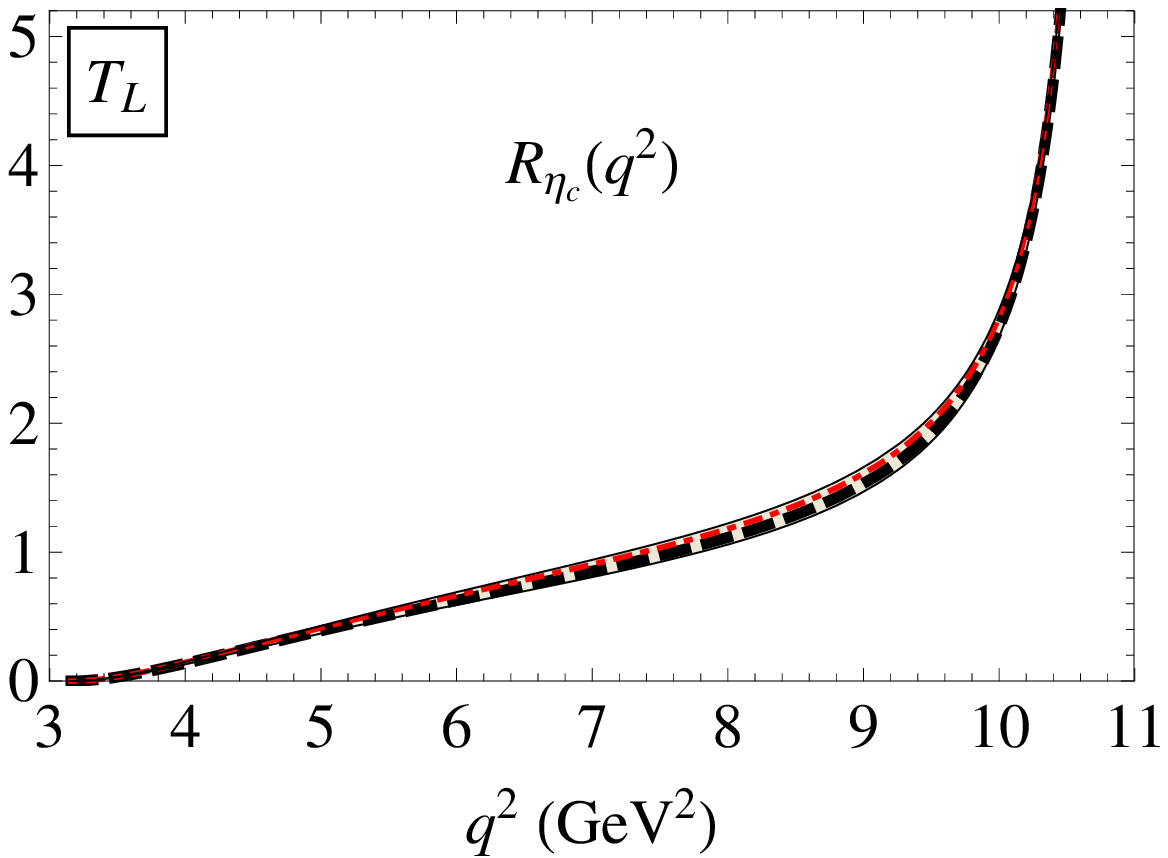}&
\includegraphics[scale=0.5]{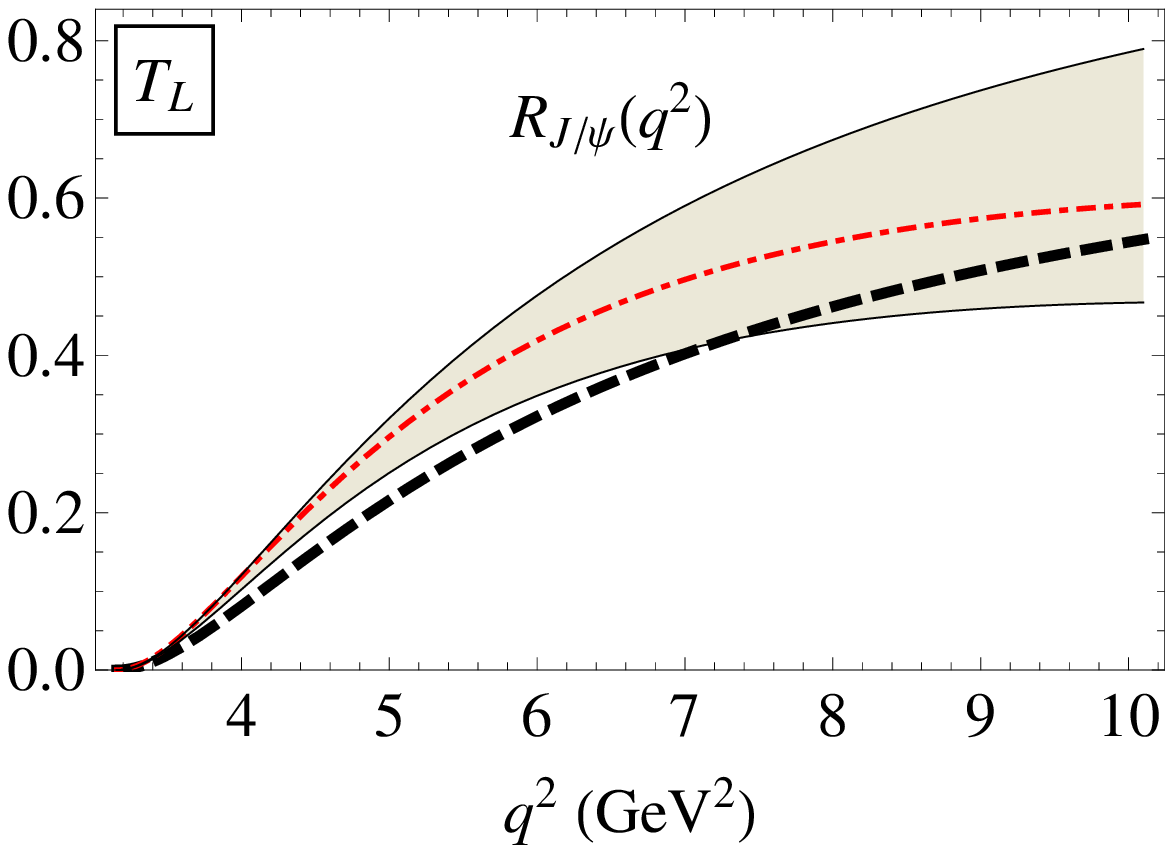}
\end{tabular}
\caption{Differential ratios $R_{\eta_c}(q^2)$ (\textbf{left}) and $R_{J/\psi}(q^2)$ (\textbf{right}) in the tensor scenario. The black dashed lines are the SM predictions; the gray bands include NP effects corresponding to the $2\sigma$ allowed regions in Figure~\ref{fig:VT2sigma}; the red dot-dashed lines represent the best-fit value of the corresponding NP coupling given in Eq.~(\ref{eq:bestfit}).}
\label{fig:R}
\end{figure}

In this paper, we redo the analysis using the most updated data, and we present here only significant changes compared with our previous results. The most important update is that the tensor coupling allowed at $2\sigma$ has a negligible effect on the ratios $R_D$ and $R_{\eta_c}$ in comparison with that on $R_{D^*}$ and $R_{J/\psi}$. For demonstration, we show in Figure~\ref{fig:R} the differential ratios $R_{\eta_c}(q^2)$ and $R_{J/\psi}(q^2)$ assuming the tensor scenario. In the case of the vector operators ${\cal O}_{V_L}$ and ${\cal O}_{V_R}$, the effects on all of the ratios $R_{D^{(*)}}$ and $R_{J/\psi(\eta_c)}$ are rather similar. 
%%%%%%
\unskip
In Table~\ref{tab:R} we present the average values of $R_{\eta_c}$ and $R_{J/\psi}$ over the whole $q^2$ region. The predicted ranges for the ratios in the presence of NP are given in correspondence with the $2\sigma$ allowed regions of the NP couplings shown in Figure~\ref{fig:VT2sigma}.
%%%%
\begin{table}[htbp] 
\centering
\caption{The $q^2$ average of the ratios in the Standard Model and in the presence of New Physics.}
\begin{tabular}{c|cc}
\hline\hline
\toprule
&\quad  {\boldmath $<R_{\eta_c}>$} \qquad 
&\quad  {\boldmath $<R_{J/\psi}>$} \qquad   
\\
\midrule
\hline
{\textbf {SM}} &\quad $0.26$\quad &\quad $0.24$\quad \\
{\boldmath $V_L$}
&\quad $[0.26,0.38]$\quad
&\quad $[0.25,0.35]$\quad
\\
{\boldmath $V_R$}
&\quad $[0.25,0.41]$\quad
&\quad $[0.25,0.36]$\quad
\\
{\boldmath $T_L$}
&\quad $[0.25,0.28]$\quad
&\quad $[0.24,0.36]$\quad
\\
\bottomrule
\hline\hline
\end{tabular}
\label{tab:R}
\end{table}

Finally, we focus on the prediction for the longitudinal polarization of the final tau, since it has been measured recently (in the decay $B \to D^{\ast} \tau\nu_\tau$~\cite{Hirose:2016wfn}), and more precise results are expected to be coming in the near future. The longitudinal polarization reads~\cite{Ivanov:2017mrj}
\bea
\label{eq:PL}
P_L^{P\to P^\prime}(q^2)&=&\frac{1}{{\cal H}_{\rm tot}^{P^\prime}}\Big\lbrace
-|1+g_V|^2\big[|H_0|^2-\delta_\tau(|H_0|^2+3|H_t|^2)\big]+3\sqrt{2\delta_\tau}{\rm Re}g_S H_P^S H_t\nn
&&+\frac{3}{2}|g_S|^2|H_P^S|^2
+8|T_L|^2(1-4\delta_\tau)|H_T|^2-4\sqrt{2\delta_\tau}{\rm Re}T_L H_0 H_T
\Big\rbrace,\\
P_L^{P\to V}(q^2)&=&\frac{1}{{\cal H}_{\rm tot}^{V}}\Big\lbrace
(|1+V_L|^2+|V_R|^2)\big[-\sum\limits_{n}|H_{n}|^2+\delta_\tau(\sum\limits_{n}|H_{n}|^2+3|H_{t}|^2)\big]\nn
&&-2{\rm Re}V_R\big[(1-\delta_\tau)(-|H_{0}|^2+2H_{+}H_{-})+3\delta_\tau |H_{t}|^2\big]-3\sqrt{2\delta_\tau}{\rm Re}g_P H_V^S H_{t}\\
&&+\frac{3}{2}|g_P|^2|H_V^S|^2+8|T_L|^2(1-4\delta_\tau)\sum\limits_{n=0,\pm}|H_T^n|^2+4\sqrt{2\delta_\tau}{\rm Re}T_L\sum\limits_{n=0,\pm}H_{n}H_T^n
\Big\rbrace
.
\nonumber
\ena
Note that the tau longitudinal polarization is defined in
the $W^*$ rest frame, not in the parent $B$-meson rest frame.

\begin{figure}[htbp]
\centering
\begin{tabular}{cc}
\includegraphics[scale=0.5]{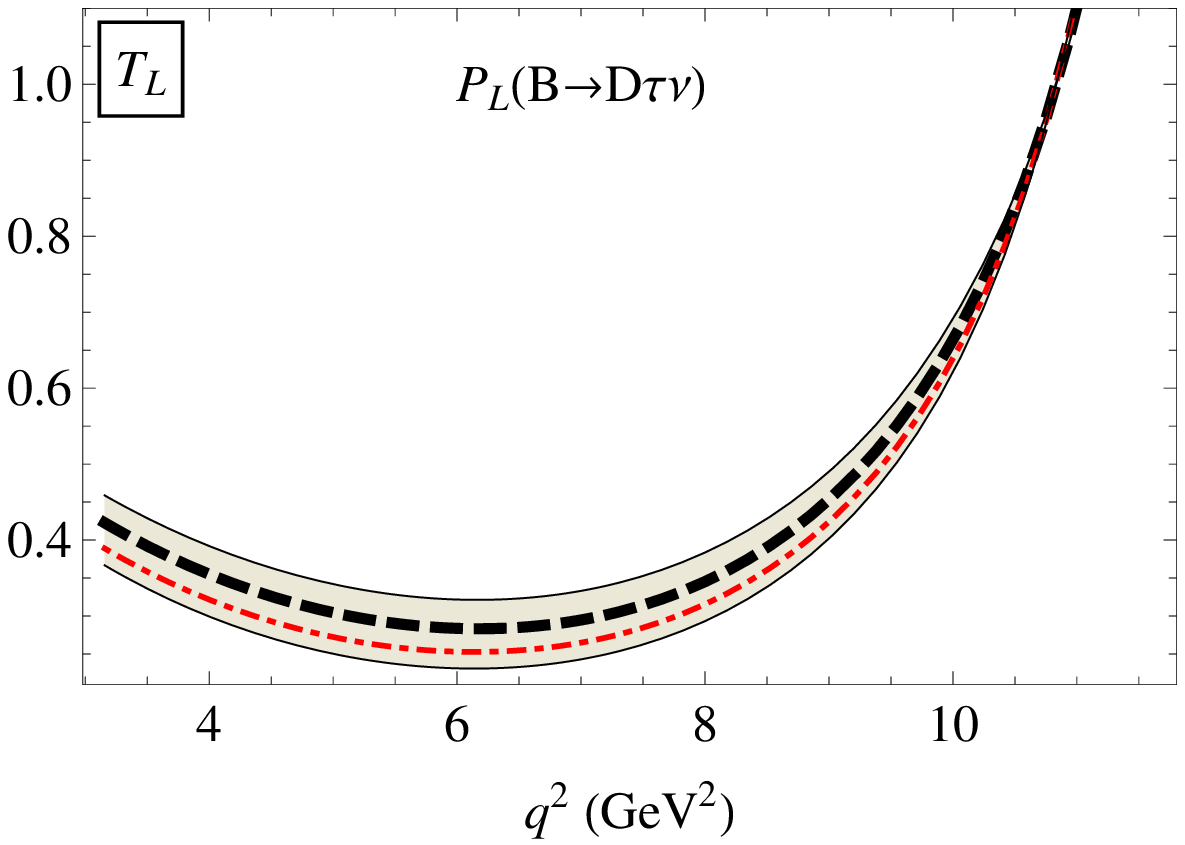}&
\includegraphics[scale=0.5]{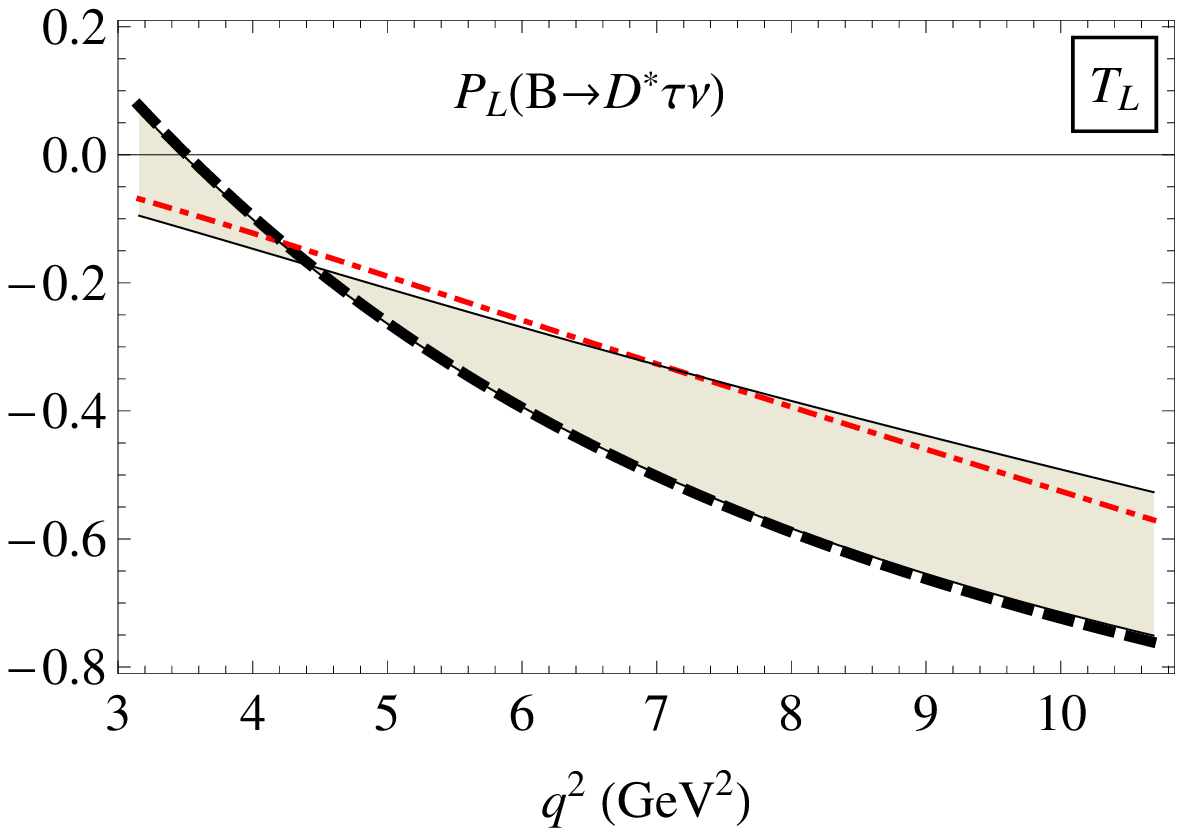}
\end{tabular}
\caption{Longitudinal polarization of the $\tau$ in the decays $B^0 \to D\tau\nu$ (\textbf{left}) and $B^0 \to D^*\tau\nu$ (\textbf{right}) in the tensor scenario. Notations are the same as in Figure~\ref{fig:R}. }
\label{fig:PL}
\end{figure}

The longitudinal polarization $P_L^\tau$ is not affected by the vector operators. The $q^2$ dependence of $P_L^\tau$ in the tensor scenario is presented in Figure~\ref{fig:PL}. The $q^2$ averaged values of $P_L^\tau$ are shown in Table~\ref{tab:PL}. The current status of experimental data has ruled out the scalar operators at $2\sigma$, and the tensor operator is the only NP contribution that has an impact on $P_L^\tau$. Moreover, the effects of ${\cal O}_{T_L}$ on $<P_L^\tau(D)>$ and $<P_L^\tau(\eta_c)>$ are small. Specifically, the best-fit value suggests a large enhancement of $<P_L^\tau(D^*)>$, which can be tested in high-precision experiments at Belle~II.

\unskip
\begin{table}[htbp] 
\centering
\caption{The $q^2$ average of the longitudinal polarization in the SM and in the presence of NP.}
\begin{tabular}{c|cccc}
\toprule
\hline\hline
&\quad  {\boldmath $<P_L^\tau(D)>$} \qquad 
&\quad  {\boldmath $<P_L^\tau(D^*)>$} \qquad   
&\quad  {\boldmath $<P_L^\tau(\eta_c)>$} \qquad 
&\quad  {\boldmath $<P_L^\tau(J/\psi)>$} \qquad   
\\
\midrule
\hline
{\textbf SM} &\quad $0.33$\quad &\quad $-0.50$\quad
&\quad $0.36$\quad &\quad $-0.51$\quad \\
{\boldmath $T_L$}
&\quad $[0.28,0.37]$\quad
&\quad $[-0.50,-0.33]$\quad
&\quad $[0.31,0.40]$\quad
&\quad $[-0.51,-0.34]$\quad
\\
\textbf{Best-fit} {\boldmath $(T_L)$} &\quad $0.30$\quad &\quad $-0.33$\quad
&\quad $0.33$\quad &\quad $-0.34$\quad
\\
\textbf{Experiment}&\quad {}\quad &\quad $-0.38 \pm 0.51^{+0.21}_{-0.16}$~\cite{Hirose:2016wfn}\quad
&\quad {}\quad &\quad {}\quad
\\
\bottomrule
\hline\hline
\end{tabular}
\label{tab:PL}
\end{table}
%%%%%%

%%%%%%%%%%%%%%%%%%%%%%%%%%%%%%%%%%%%%%%%%%
\section{Conclusions}
\label{sec:sum}

Inspired by recent Belle measurements of the polarization observables $P_L^\tau$ and $F_L^{D^*}$ in the $B\to D^*\tau\nu_\tau$ decay, as well as the ratios $R_{D^{(*)}}$, we have revisited the flavor anomalies in the semileptonic transition $b\to c\tau\nu$ based on an effective Hamiltonian consisting of vector, scalar, and tensor four-fermion operators. The form factors parametrizing the corresponding hadronic transitions $B\to D^{(*)}$ and $B_c\to J/\psi(\eta_c)$ have been calculated in our covariant constituent quark model.  Under the assumption of one-operator dominance, we have obtained the available regions for the Wilson coefficients characterizing the NP contributions using the most updated experimental constraints from the ratios $R_{D^{(\ast)}}$ and $R_{J/\psi}$, the leptonic branching $\mathcal{B}(B_c\to\tau\nu)\leq 10\%$, and the polarization fraction $F_L^{D^*}$. In particular, we have discussed the effects of the new constraint from $F_L^{D^*}$ on the overall picture. 

It turned out that at the level of $2\sigma$, the scalar coefficients $S_{L,R}$ are excluded (mainly by the constraint $\mathcal{B}(B_c\to\tau\nu)\leq 10\%$), while the vector ($V_{L,R}$) and tensor ($T_L$) ones are still available. If the upper limit on $\mathcal{B}(B_c\to\tau\nu)$ is relaxed up to 30\%, then $S_L$ and $S_R$ are also allowed at $2\sigma$, but only minimally. However, all coefficients are ruled out at $1\sigma$. The recent measurement of $F_L^{D^*}$ provides a severe constraint on the tensor scenario. In particular, the tensor scenario is excluded at $1\sigma$ by the constraint from $F_L^{D^*}$ alone. We have also observed that the effects of the tensor operator on the differential ratios $R_D(q^2)$ and $R_{\eta_c}(q^2)$ are now negligible. Finally, within the $2\sigma$ allowed regions of the corresponding Wilson coefficients, we have provided predictions for the $q^2$ average of the ratios of branching fractions and the tau longitudinal polarization, which will be useful for future testing of LFU in these decays.

\begin{acknowledgments}
M.A.I. acknowledges support from PRISMA$^+$ Cluster of Excellence (Mainz Uni.). M.A.I. and J.G.K. acknowledge support from Heisenberg-Landau Grant for their collaboration. P.S. acknowledges support from Istituto Nazionale di Fisica Nucleare, I.S. QFT\_HEP. 
J.G.K., P.S., and C.T.T. would like to thank the organizers of the 2019 Helmholtz International Summer School (HISS)--Dubna International Advanced School of Theoretical Physics (DIAS-TH) “Quantum Field Theory at the Limits: From Strong Fields to Heavy Quarks” for their support, as well as the warm hospitality during the school. C.T.T. thanks Aidos~Issadykov and Vu~Duc~Cong for their heartiness and interesting discussions.
\end{acknowledgments}

%%%%%%%%%%%%%%%%%%%%%%%%%%%%%%%%%%%%%%%%%%
\end{document}